\documentclass[aas_macros]{aa}  
\usepackage{txfonts}
\usepackage{amsmath}
\usepackage{amsfonts}
\usepackage{amssymb}
\usepackage{gensymb}
\usepackage{color}
\RequirePackage[colorlinks, citecolor=blue, urlcolor=magenta, linkcolor=darkblue, bookmarks=false, breaklinks=true]{hyperref}
\usepackage[]{natbib}
\bibpunct{(}{)}{;}{a}{}{,}
\definecolor{darkblue}{rgb}{0.0, 0.0, 0.62}
\definecolor{deepmagenta}{rgb}{0.8, 0.0, 0.5}
\definecolor{darkred}{rgb}{0.55, 0.0, 0.0}
\definecolor{violet}{rgb}{0.6, 0.0, 1.0}

\RequirePackage{mymacro}

\graphicspath{{./fig/}}

\begin{document} 

\title{Untangling dust emission and cosmic infrared background anisotropies with the scattering transform statistics}

\titlerunning{Untangling dust emission and cosmic infrared background anisotropies}
   
\author{Srijita Sinha \thanks{Corresponding author: sinha.srijita@niser.ac.in} \inst{1,2}, Tuhin Ghosh \thanks{Corresponding author: tghosh@niser.ac.in} \inst{1,2},
Erwan Allys\inst{3}, Fran\c{c}ois Boulanger\inst{3},
% \and
Jean-Marc Delouis\inst{4}
}

\institute{School for Physical Sciences, National Institute of Science Education and Research, HBNI Jatni-752050, India
\and
Homi Bhabha National Institute, Training School Complex, Anushakti Nagar, Mumbai 400094, India
\and
Laboratoire de Physique de l'{\'E}cole Normale Sup{\'e}rieure, ENS, Université PSL, CNRS, Sorbonne Universit{\'e}, Universit{\'e} Paris Cit{\'e},
75005 Paris, France
\and
Laboratoire d'Oc{\'e}anographie Physique et Spatiale (LOPS), Univ. Brest, CNRS, Ifremer, IRD, 29200 Brest, France
}
\authorrunning{S.\ Sinha et al.}
             
\date{} 

%%%%%%%%%%%%%%%%%%%%%%%%%%%%%%%%%%%%%%%%%%%%%%%%%%%%%%
  \abstract
   {A template-fit approach is often used to separate the Galactic dust emission and the cosmic infrared background (CIB) anisotropies in low \ion{H}{I} column density regions using the observational fact that the $21\,\rm cm$ \ion{H}{I} line emission from neutral atomic hydrogen and dust are tightly correlated. However, in some regions with molecular hydrogen, diffuse ionised gas, and dark gas, the same approach fails to trace the excess Galactic dust emission.}
   {We developed and tested a statistical component-separation method to extract the dust signal from the contaminated \textit{Planck} $353\,\rm GHz$ observations using the scattering covariance (SC) statistics, which is a subclass of scattering transform statistics.}
   {We first obtained a set CIB maps over $25$ square patches, each with a sky area of $222\,\deg^{2}$, using the linear correlation of dust and Galactic \ion{H}{I} column density map valid in low \ion{H}{I} column density regions using the template-fit approach. We then constructed from these $25$ maps a generative model of CIB using SC statistics. We finally relied on this generative model to perform a component-separation of dust and CIB in the \textit{Planck} data for different sky regions. These separations were achieved by sampling an ensemble of dust maps through pixel-based optimisation, which, when added to the CIB contamination model, verified all the statistics and cross statistics constraints that were estimated directly from the data.}
   {We validated our algorithm and separated the dust emission from the contamination in the \textit{Planck} $353\, \rm GHz$ observations. We show the results of the recovered dust map for a test sky region where there is a significant difference between the \textit{Planck} dust map and CSFD map. We found that the \textit{Planck} dust map has more structure than the CSFD map. We compared the power spectrum of the recovered dust map and \ion{H}{I} map and found a difference in slope ($\Delta \alpha \approx 0.4$) by fitting a power-law model to the two spectra. To explain $\Delta \alpha$, we decomposed the recovered dust map into two gas phases: dust associated with neutral atomic hydrogen, and dust associated with molecular hydrogen. We provide a clear pathway to mapping the Galactic interstellar reddening over intermediate and high Galactic latitudes. 
    } 
   {}
   \maketitle

%%%%%%%%%%%%%%%%%%%%% BODY OF PAPER %%%%%%%%%%%%%%%%%%%%%
\section{Introduction} \label{sec:intro}

The cosmic microwave background (CMB) signal encoding the history of the Universe at microwave and far-infrared frequencies is buried deep below the large-scale emission from the Galaxy and the small-scale emission from the extra-galactic sources~\citep{ichiki2014ptep}. At frequencies above  $200\GHz$, diffuse thermal emission from our own Galaxy and cosmic infrared background (CIB;~\citealt{Puget:1996, Lagache:2005sw}) are the primary contributors to foreground emission~\citep{planck-IV:2018}. Galactic thermal dust emission provides a wealth of information regarding the complex physical systems that form the interstellar medium (ISM)~\citep{draine2011book}. The CIB is the diffuse radiation from dust particles in star-forming galaxies through the evolution of the Universe, and it acts as a probe for the dark matter distribution and star formation history~\citep{hauser2001araa, planck-XXX:2014, maniyar2018aa}. Moreover, the CIB plays a vital role in de-lensing the observed CMB $B$ modes~\citep{Larsen:2016}. Although they have very different origins, dust emission and the CIB share a similar spectral energy distribution; both follow a modified blackbody spectrum with a slightly different spectral index. Separating these two components is challenging using spectral information alone. 

Galactic thermal dust emission was observed to be tightly correlated with the $21\,\cm$ \HI emission in low \HI column density regions~\citep{Boulanger:1996, Boulanger:1988}. For this reason, the \HI emission map is often used as a tracer of the dust emission within a template-fit approach to separate it from the CIB anisotropies~\citep{planck-XVII:2014, adak2024mnras} and study its spectral energy distribution. Several other attempts have been made to obtain a clean map of CIB anisotropies by taking advantage of the full velocity coverage  ($|v_{LSR}| < 600 \kms$) of $21\,\cm$ \HI observations~\citep{planck-XXIV:2011, Lenz:2019, Fiona:2024}, where $v_{LSR}$ is the velocity of the \HI gas measured in the local standard of rest frame. The dust-CIB separation using the template-fit approach, however, becomes challenging when the zero level of the sky background is unknown and there are significant variations in the dust emissivity at high Galactic latitudes.~\cite{planck-VIII:2013} used a constant dust emissivity in very  low \HI column density regions ($\NHI < 2 \times \hiunit$) to set the zero level of the \Planck intensity map at $857 \GHz$. \Planck $857 \GHz$ was then used as a reference to set the zero level of all other \Planck high frequency instrument (HFI) intensity maps. More recently,~\cite{adak2024mnras} generalised this approach by fitting a pixel-dependent dust emissivity and a global offset using the Galactic \HI template within a Bayesian inference framework. The dust and CIB can be separated at the power spectrum level, where the \HI power spectrum is used as a tracer of the dust power spectrum and subtracted from the total map to estimate the CIB power spectrum~\citep{Mak:2017,viero2019apj}. Another model-independent approach used the generalized needlet internal linear combination method, which employs spatial information of the CIB in terms of its angular power spectrum to disentangle dust emission from CIB anisotropies~\citep{planck-XLVIII:2016}.  

Another approach to separate dust emission from CIB anisotropies is to rely on scattering transform (ST) statistics. ST statistics are a family of low-variance summary statistics inspired by neural networks that efficiently characterise non-Gaussian processes~\citep{bruna2013ieee, anden2014ieee}. ST statistics are constructed from convolutions with a set of pass-band wavelets that separate a signal into its different oriented scales, and non-linear operations, such as modulus, that characterise the interaction, that is, the statistical dependence, between these different scales~\citep{cheng2021ax}. Since their introduction in astrophysics, these statistics have consistently approached or reached the best possible performance in characterising, classifying, and inferring parameters in various domains, such as the interstellar medium~\citep{allys2019aa, regaldo2020aa, saydjari2021apj, lei2023apj, richard2025aa}, the large-scale structures of the Universe~\citep{allys2020prd, eickenberg2022ax, valogiannis2022prd}, weak lensing~\citep{cheng2020mnras, cheng2021mnras}, and the epoch of reionisation~\citep{greig2024mnras, hothi2024aa}. 

Additionally, STs offer the ability to construct an approximate generative model of a given process from an estimate of its ST statistics~\citep{bruna2018jmst}. This has been demonstrated for various physical processes, with models sometimes constructed from a single image~\citep{allys2020prd,cheng2024pnas}. Initially developed for two-dimensional (2D) planar data, these models have since been expanded to include multi-frequency data~\citep{regaldo2023apj}, spherical data~\citep{mousset2024aa, campeti2025aa}, and even spectroscopic data~\citep{hothi2025generative}. Simulated data in CMB studies showed that an ST generative model constructed from a single dust-foreground patch can suffice for training a neural network to distinguish primordial $B$ modes from dust emission, even within a challenging mono-frequency approach~\citep{jeffrey2021mnrasl}.

Moreover, new ST-based component-separation algorithms have been developed. Introduced in~\cite{regaldo2021aa, delouis2022aa} to separate Galactic dust emission from instrumental noise in \Planck data, these versatile algorithms relied on pixel-space optimisation under ST statistics constraints to generate dust maps compatible with the data once added to the noise. In these papers, the ability of these algorithms to leverage the different non-Gaussian properties of different signals was emphasised by the fact that the separations were performed at a single frequency. Additionally, these separations were performed without assuming a prior model of the component of interest (dust, in this case). Recently,~\cite{auclair2024aa} used an ST-based algorithm to separate dust emission from CIB anisotropies in \herschel Spectral and Photometric Imaging Receiver (SPIRE) observations. A major success of this paper is that this separation was achieved by relying on observational data alone, first learning an ST-based CIB model from a sky region where this signal was dominant, and then using this model to separate Galactic dust emission and CIB in the Spider region. More recently, \cite{tsouros2026ar} used the ST-based component-separation to recover the polarised dust emission from the \Planck data at $353 \GHz$. 

The main goal of this paper is to build on previous work and separate the Galactic dust emission in \Planck $353 \GHz$ data from the CIB anisotropies and instrumental noise. To do this, we first construct an ST-based contamination model of CIB and instrumental noise over a set of square patches, each with a sky area of $222\,\deg^{2}$, in regions of low \HI column density. Secondly, we use this contamination model to perform an ST-based component-separation in regions of intermediate \HI column densities. We use the particular scattering covariance (SC) statistics, a sub-class of the ST statistics. We work with \Planck $353 \GHz$ data smoothed to a angular beam resolution of $16.2\arcm$ (full width at half maximum; FWHM) to match the angular beam resolution of the \HI emission maps. At this angular beam resolution, the CIB anisotropies dominate instrumental noise. Before applying the component-separation algorithm to the real \Planck data, we successfully test it on the simulated \Planck maps with different signal-to-noise ($\mathrm{S/N}$) ratios  of the dust emission with respect to the contamination. 

The paper is organised as follows. In Sect.\ \ref{sec:data-set} we describe the \Planck data and the external datasets (\texttt{HI4PI} and dust-reddening map at 100\,$\mu$m). Section \ref{sec:hmc-cib} briefly describes the results of the template-fit approach to obtain the contamination map (or dust-subtracted \Planck map) at $353 \GHz$ over the low \HI column density regions. In Sect.\ \ref{sec:cib-ps-all} we estimate the statistical properties of the contamination map from a limited set of square patches and its sample variance. In Sect.\ \ref{app:scat-coeff} we briefly present the set of ST statistics. The component-separation algorithm using the ST statistics is described in Sect.\ \ref{sec:wph-dust-cib}. In Sect.\ \ref{sec:planck-simulation} we apply our algorithm to \Planck simulations at $353 \GHz$. We present and discuss the results of the dust-CIB separation obtained from the \Planck $353 \GHz$ observations in Sect.\ \ref{sec:results}, and finally, we summarise our findings in Sect.\ \ref{sec:discussion}.
%%%%%%%%%%%%%%%%%%%%%%%%%%%%%%%%%%%%%%%%%%%%%%%%%%%%%
\section{Datasets}\label{sec:data-set}

\subsection{Planck data}\label{sec:planck-data}

We used the publicly available \Planck\footnote{\href{http://pla.esac.esa.int/pla}{http://pla.esac.esa.int/pla}} spectral matching independent component analysis (\texttt{SMICA}) CMB-subtracted intensity map from the Public Release 3 at $353 \GHz$~\citep{planck-I:2018}. The $353 \GHz$ map is provided in \healpix\footnote{\href{http://healpix.sf.net}{http://healpix.sf.net}}~\citep{Gorski:2005} format at $\Nside=2048$ (pixel size $\sim 1.7\arcm$) with an angular beam resolution of $4.82\arcm$ FWHM~\citep{planck-III:2018, planck-IV:2018}. We then smoothed the map at an angular beam resolution of 16.2\arcm\ (by convolving it with an additional Gaussian beam of $\sqrt{(16.2\arcm)^2 - (4.82\arcm)^2}=15.47\arcm$) and downgraded it to $\Nside=512$ (pixel size $6.8\arcm$). We converted the map from $\kcmb$ to $\kJysr$ units using the conversion factors given in~\citet{planck-IX:2014}. We retained the CIB monopole that was added by hand to the \Planck data at $353 \GHz$. From the \healpix map, we extracted 2D tangential projection square patches ($256\times 256$ pixels) centred around \healpix pixel of $\Nside=4$ using the \texttt{reproject} Python package~\citep{robitaille2020}. The pixel size for each square patch was $3.5 \arcm$, resulting in a total patch area of $222\,\deg^{2}$.

We used the end-to-end full focal plane 10 (FFP10) noise simulations at $353 \GHz$~\citep{planck-III:2018} to estimate the variance of the instrumental noise in the \Planck $353 \GHz$ data. Similar to the \Planck data, we first smoothed the noise maps with an additional Gaussian beam smoothing of $15.47\arcm$, reprojected them to $\Nside=512$, and then extracted the 2D patches from them. 
%%%%%%%%%%%%%%%%%%%%%%%%%%%%%%%%%%%%%%%%%%%%%%%%%%%%%
\subsection{External datasets}\label{sec:hi-map}

We used the \texttt{HI4PI} full-sky map with a spectral resolution of $1.49\,\kms$, which combines the data from the Effelsberg-Bonn \HI Survey (EBHIS;~\citealt{ebhis2016aa}) and the Galactic All-Sky Survey (GASS;~\citealt{GASSI:2009, GASSII:2010}). The map is provided on the \healpix grid at $\Nside=1024$ (pixel size $3.4\arcm$) and a common angular beam resolution of $16.2\arcm$ (FWHM). The root mean square brightness temperature uncertainty is $43\,\rm mK$. Along the optically thin line of sight, the total \HI column density (\NHI; in units of $10^{18}\,\cm^{-2}$) can be obtained by integrating the brightness temperature ($T_b$) over the velocity channels  using the relation~\citep{Dickey:1990}\begin{equation}
\NHI = 1.82 \int \, T_b \, \mathrm{d}v_{LSR} \,. 
\end{equation}
Following~\cite{hayakawa2024mnras}, we integrated $T_b$ over two velocity ranges of the \HI clouds: one range with a low velocity (LV; $\abs{v_{LSR}} < 30\,\kms$), and the other with an intermediate velocity (IV; $30\,\kms < \abs{v_{LSR}} < 100\,\kms$).
We ignored the high-velocity \HI emission (HV; $|v_{LSR}| > 90\,\kms$) as no significant dust emission is associated with the HV template~\citep{Wakker:1986, planck-XXIV:2011, Lenz2016, hayakawa2024mnras}. The column densities associated with the LV and IV components are defined as \NLV and \NIV, respectively. Finally, we downgraded the \NLV and \NIV maps to the \healpix grid of $\Nside=512$ and extracted 2D square patches centred around \healpix pixel of $\Nside=4$. We treated the two column density maps as the tracers of the dust emission. We used the sum of the \NIV and \NLV as a measure of \NHI. 

The all-sky Galactic dust-reddening map produced by \citet*[][hereafter SFD]{SFD:1998} at $100\,\mum$ has imprints of extragalactic large-scale structure (LSS) or CIB \citep{chiang2019apj}. The CIB map at $100\,\mum$ ($\rlss$) was reconstructed by cross-correlating the SFD map with the spectroscopic galaxies and quasars in SDSS. The corrected SFD (CSFD) was produced by subtracting the CIB contamination from the SFD map\footnote{\href{https://doi.org/10.5281/zenodo.8207175}{https://doi.org/10.5281/zenodo.8207175}} \citep{chiang2023apj}. We worked with a publicly available CSFD map and a $100\,\mum$ CIB map at a \healpix resolution of $\Nside=512$ and smoothed the two maps to a common angular beam resolution of $16.2\arcm$ (FWHM). The CSFD map was used as a tracer for the dust emission to produce the \Planck simulations.
%%%%%%%%%%%%%%%%%%%%%%%%%%%%%%%%%%%%%%%%%%%%%%%%%%%%%
\section{Contamination map from the template-fit approach}\label{sec:hmc-cib}

In this section, we briefly discuss how we obtained the contamination map (or dust-subtracted \Planck map) at $353 \GHz$. We used the \cite{adak2024mnras} formalism to fit the pixel-dependent dust emissivity and a global offset over low \NHI regions (${\NHI < 4 \times \hiunit}$) using the template-fit approach. We used LV and IV \HI templates as a tracer for the dust emission. Because the LV and IV maps are full-sky maps, we easily included the northern and southern Galactic hemispheres in the analysis, and we hence increased the total sky coverage.

We followed the iterative correlation method of~\citet{planck-XVII:2014} to construct a global mask, which is defined as follows. Over the initial mask with ${\NHI\,<\,4\,\times\,\hiunit}$, we computed the pixel-dependent dust emissivity between the CMB-subtracted \Planck intensity map at $353 \GHz$ and the two \HI templates, and we then subtracted the best-fit model from the \Planck $353 \GHz$ data to produce a residual map. We fit a Gaussian to the residuals, calculated its standard deviation ($\sigma_{\rm G}$), and removed pixels with absolute values greater than $5\sigma_{\rm G}$. We repeated the same procedure until all pixels in the residual map fell within $5\sigma_{\rm G}$ region of the Gaussian fit to the residuals. We arrived at a converged Galactic mask after five iterations, covering a region of $\sim 13800 \deg^{2}$. The total sky fraction covered by the unmasked pixels is $\fsky = 0.33$. 

We modelled the input CMB-subtracted \Planck intensity data ($m$) at $353 \GHz$ as the sum of the dust emission ($s$), CIB anisotropies ($c$), and the instrumental noise ($n$),
\begin{equation}
    m =  s + c + n.
\end{equation}
The global offset of the map (including the CIB monopole) term was included in the signal term $s$. Using~\cite{adak2024mnras} formalism, we sampled the joint probability distribution of the pixel-dependent dust emissivity and the global offset given the two \HI templates and the input data. We modelled $s$ as
\begin{equation}
     s = \epsilon_{\rm LV} \NLV + \epsilon_{\rm IV} \NIV + O \ ,
 \end{equation}
where $\epsilon_{\rm LV}$ and $\epsilon_{\rm IV}$ correspond to the dust emissivity associated with the LV and IV column density maps, respectively, and $O$ is the global offset. We assumed that the two dust emissivities varied over the sky, but had fixed values within a $1.8\degree \times 1.8\degree$ pixel area, corresponding to a single pixel area of a \healpix $\Nside=32$ map. The contributions of CIB anisotropies and instrumental noise were propagated through the noise covariance matrix~\citep{adak2024mnras}. 
%%%%%%%%%%%%%%%%%%%%%%%%%%%%%%%%%%%%%%%%%%%%%%%%%%%%%
%%%%%%%%%%%%%%%%%%%%%%%%%%%%%%%%%%%%%%%%%%%%%%%%%%%%%
The emissivity maps are shown in Fig.\ \ref{fig:emissivity-maps}. The mean and $1\sigma$ standard deviation values of $\epsilon_{\rm LV}$ and $\epsilon_{\rm IV}$ of 2644 $\Nside=32$ pixels  over the northern Galactic hemisphere are $46.5 \pm 13.3$ and $9.2 \pm 41.2\, \kJysr\paren{\hiunit}^{-1}$, respectively. The mean values of $\epsilon_{\rm LV}$ and $\epsilon_{\rm IV}$ of $2257$ $\Nside=32$ pixels over the southern Galactic hemisphere are $40.5 \pm 10.3$ and $-17.2 \pm 32.7 \, \kJysr\paren{\hiunit}^{-1}$, respectively. 
We report a global offset value of $O=119.3\pm0.2\,\kJysr$ over the Galactic mask used in our analysis. Our value is very close to the CIB monopole of $130\,\kJysr$ added to the \Planck map at $353 \GHz$ based on the~\cite{Bethermin:2010} CIB model. The small difference between our estimate of the global offset and the CIB monopole term added by the \Planck collaboration of roughly $10.7\,\kJysr$ (or $37.5\,\mu\kcmb$) might indicate warm ionized medium associated dust emission at high Galactic latitude \citep{gaensler2008pasa}, as included in the \Planck analysis \citep{planck-XII:2018}.
%%%%%%%%%%%%%%%%%%%%%%%%%%%%%%%%%%%%%%%%%%%%%%%%%%%%%
%%%%%%%%%%%%%%%%%%%%%%%%%%%%%%%%%%%%%%%%%%%%%%%%%%%%%
%%%%%%%%%%%%%%%%%%%%%%%%%%%%%%%%%%%%%%%%%%%%%%%%%%%%%
\begin{figure}[!htbp]
\centering
\includegraphics[width=\columnwidth]{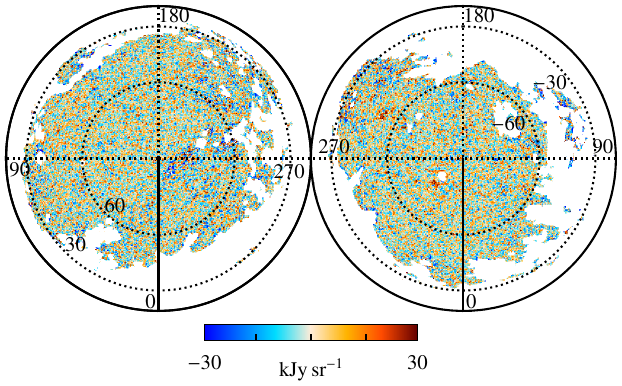}
\caption{Orthographic projection of the residual map obtained from the template-fit approach. The northern (southern) Galactic hemisphere is shown on the left (right). The white regions are masked from the analysis as they comprise \NHI cut-off pixels and the pixels were masked due to iterative masking.}
\label{fig:hmc-cib-full-sky}
\end{figure}
%%%%%%%%%%%%%%%%%%%%%%%%%%%%%%%%%%%%%%%%%%%%%%%%%%%%%%
%%%%%%%%%%%%%%%%%%%%%%%%%%%%%%%%%%%%%%%%%%%%%%%%%%%%%%

We refer to the dust map estimated from the template-fit approach within the Bayesian inference framework as $\sB$. Next, we subtracted $\sB$ from the input CMB-subtracted map to obtain the residual map, $\rB= m - \sB$, which includes the CIB anisotropies, instrumental noise, and possible residual dust emissions (not correlated with the two \HI templates). The expected standard deviation of instrumental noise from FFP10 simulations over the Galactic mask is $3.3\,\kJysr$, whereas the expected standard deviation of the CIB anisotropies based on the \Planck 2013 best-fit CIB model~\citep{planck-XXX:2014} is $9.3\, \kJysr$. Because the standard deviation of two uncorrelated components is added in quadrature, we safely assumed that $\rB$ map is dominated by the CIB anisotropies. The residual map at $353 \GHz$ over the Galactic mask is shown in Fig.\ \ref{fig:hmc-cib-full-sky} in orthographic projection. With the two-template fit, we were able to extract the \HI-correlated dust model map and the residual map over roughly twice the sky fraction as compared to the one-template fit analysis in the southern Galactic cap region \citep{adak2024mnras}. In both cases, the one-dimensional probability distribution of the residuals closely follows the Gaussian distribution. We computed the width of the Gaussian fit ($w_{\rB}$) to the residuals over the common mask generated from the union of our iterative mask and southern Galactic cap region \citep{adak2024mnras}. We report a marginally smaller width in our analysis ($w_{\rB}=9.6$\,\kJysr) as compared to the one-template fit analysis ($w_{\rB}=9.9$\,\kJysr), indicating less leakage of Galactic emission in the residual map.
%%%%%%%%%%%%%%%%%%%%%%%%%%%%%%%%%%%%%%%%%%%%%%%%%%%%%%
\section{Statistical properties of the residual map over square patches}\label{sec:cib-ps-all}

For our purpose, we first extracted $48$ square patches ($24$ in the northern and $24$ in the southern Galactic hemisphere) of size $14.9\degree \times 14.9\degree$ (sky area of $222 \deg^2$) centred around the centre of $\Nside=4$ \healpix grid pixels at high Galactic latitudes ($\abs{b} > 45\degree$). The pixel resolution of each square patch was $3.5\arcm$ and contained $256\times 256$ pixels. We note that $\Nside=4$ pixels have a pixel size of $14.7\degree \times 14.7\degree$. All the square patches have very minimal overlap, and they hence provide an independent measurement of the statistical properties of the residual map at $353 \GHz$. We selected only $25$ of these $48$ square patches that were not severely affected by the initial mask to select the low \NHI\ regions and the pixels that were masked due to the iterative masking algorithm discussed in Sect.\ \ref{sec:hmc-cib}. Seventeen of these $25$ square patches are completely unaffected by the final mask. Only a few pixels in the remaining $8$ square patches are slightly affected by the final mask. The percentage of these masked pixels in these patches is lower than $1.5\%$. For these $25$ square patches, the $\mathrm{S/N}$ defined as $\sigma_{\sB}/\sigma_{\rB}$ varies from 1 to 2.9. 

\begin{figure}[!htbp]
\centering
\includegraphics[width=\columnwidth]{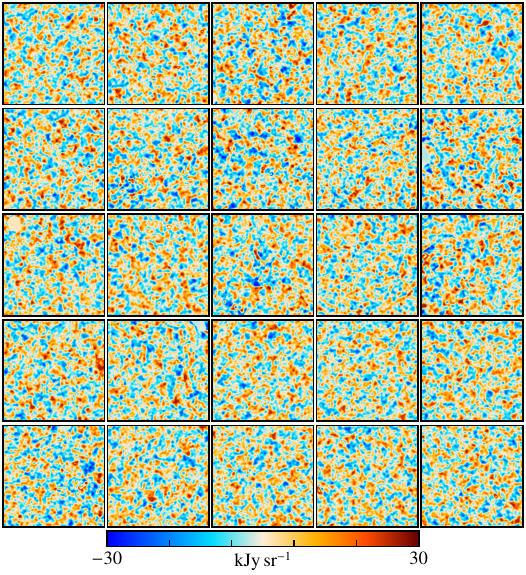}
\caption{Spatial distribution of the residual maps in the selected $25$ square patches.} 
\label{fig:hmc-cib-patches}
\end{figure}

As the template-fit approach was performed over an area of $1.8\degree \times 1.8\degree$ pixels, we expect some leakage of the Galactic dust emission from either the localised Galactic sources present in the \Planck map below the $1.8\degree$ scale or the dust emission associated with molecular hydrogen (\molH) gas or ionised hydrogen (\HII) into the residual map.  To avoid Galactic residuals in the clean sky patches, we further applied a threshold mask to the $25$ square patches. We excluded the pixels lying on the non-Gaussian tail part of the one-dimensional probability distribution function of $\rB$ map by placing a threshold at $\pm 3\sigma_{\rB}$. The percentage of the pixels that were masked due to the threshold mask ranges between $0.3\%$ to $0.8\%$. We inpainted the masked pixels using linear interpolation with \texttt{griddata} function from \texttt{scipy.interpolate} package~\citep{virtanen2020}. Finally, the inpainted patches had no pixels that deviated by $\pm 3\sigma_{\rB}$, which removed pixels containing contamination by the Galactic residuals. Figure \ref{fig:hmc-cib-patches} presents all the patches after inpainting was used to derive the statistical properties of the residual map. All the patches follow a Gaussian distribution with a mean standard deviation ($\sigma_{\rB}$) of $8.9$\,\kJysr. The estimated standard deviation of $\sigma_{\rB}$ from these patches is $0.4$\,\kJysr, which is much smaller than its mean value.
%%%%%%%%%%%%%%%%%%%%%%%%%%%%%%%%%%%%%%%%%%%%%%%%%%%%%
%%%%%%%%%%%%%%%%%%%%%%%%%%%%%%%%%%%%%%%%%%%%%%%%%%%%%
%%%%%%%%%%%%%%%%%%%%%%%%%%%%%%%%%%%%%%%%%%%%%%%%%%%%%
%%%%%%%%%%%%%%%%%%%%%%%%%%%%%%%%%%%%%%%%%%%%%%%%%%%%%
%%%%%%%%%%%%%%%%%%%%%%%%%%%%%%%%%%%%%%%%%%%%%%%%%%%%%
\begin{figure} [!htbp]
\centering
\includegraphics[width=\columnwidth]{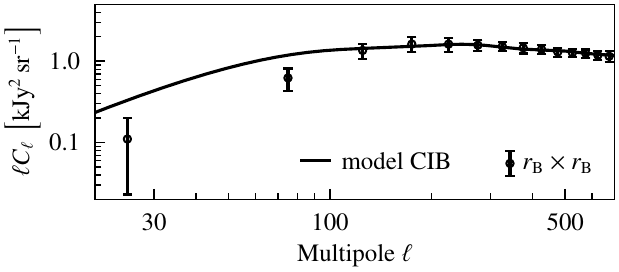}
\caption{Average power spectrum, in terms of $\ell \Cl$, with multipole $\ell$ over all the $25$ selected square patches (black points). The error bars on them are the standard deviation obtained on the spectra. The solid black line represents the best-fit CIB model by~\cite{Lenz:2019} at $353 \GHz$.} 
\label{fig:hmc-cib-ps-hist}
\end{figure}
%%%%%%%%%%%%%%%%%%%%%%%%%%%%%%%%%%%%%%%%%%%%%%%%%%%%%
We computed the power spectrum ($\Cl$) in these patches using the discrete Fourier transform in the flat-sky approximation~\citep{Hivon:2002}, implemented in the publicly available package \namaster\footnote{\href{https://github.com/LSSTDESC/NaMaster}{https://github.com/LSSTDESC/NaMaster}}\citep{alonso2019mnras}. As these patches do not obey periodic boundary conditions, we masked the boundary pixels in the computation of the power spectrum. We first made a binary mask considering only $180\times 180$ inner pixels, as shown in the left panel of Fig.\ \ref{fig:square-mask}. We then tapered the boundary with a cosine function to create the apodised mask (see the right panel of Fig.\ \ref{fig:square-mask}). The errors on the power spectrum were computed from the analytic Gaussian covariance matrix using \namaster. We corrected the final power spectrum for the beam effect and the pixel window effect.

The average power spectrum of these patches is shown in  Fig.\ \ref{fig:hmc-cib-ps-hist}. The sample variance of the power spectrum between all the patches agrees well for the spectra and is broadly consistent with the~\cite{Lenz:2019} best-fit CIB model (shown with a solid black line). At higher multipoles $\ell > 100$, our measurements of the average power spectrum of the residuals and the~\cite{Lenz:2019} CIB model match very well. 
We show the average power spectrum up to $\ell_{\rm max}=700$ because the higher multipoles are significantly affected by the beam correction.

\begin{figure}
\centering
\includegraphics[width=\columnwidth]{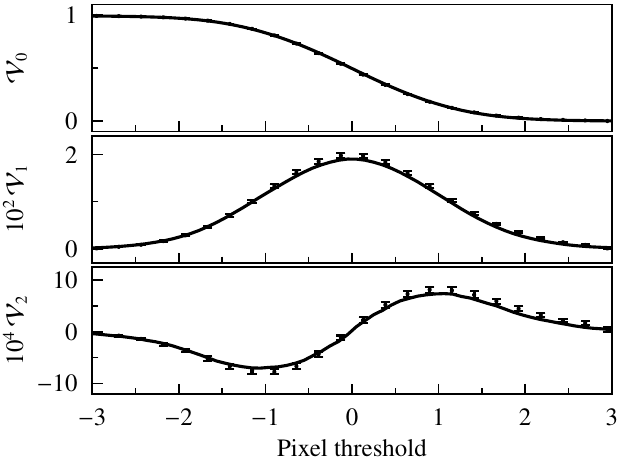}
\caption{Variation in $\mathcal{V}_{0}$, $\mathcal{V}_{1}$, and $\mathcal{V}_{2}$ with the pixel threshold of the MFs over all the $25$ patches. The perimeter and Euler functions are scaled with $10^{2}$ and $10^{4}$, respectively, for visualisation purpose. The solid black line represents the MFs computed from the average of ten FFP10 noise-contaminated Gaussian CIB realisations over a given patch.} 
\label{fig:hmc-cib-mkf}
\end{figure}

We computed the Minkowski functionals (MFs) for these patches. The MFs provide insight into the geometrical and topological structures of these regions as a variation in the pixel threshold value. For a 2D square patch, the MFs are the area ($\mathcal{V}_{0}$), perimeter ($\mathcal{V}_{1}$), and the Euler characteristic of the connected pixels and the number of holes in the same space or genus ($\mathcal{V}_{2}$). We used the publicly available package \texttt{QuantImPy} ~\citep{mantz2008jsm, boelens2021} to compute the MFs in these regions. Figure \ref{fig:hmc-cib-mkf} shows the MFs of all the $25$ selected square patches. We used the best-fit CIB model by~\cite{Lenz:2019} to generate ten Gaussian realisations of the CIB map and add ten non-Gaussian FFP10 noise realisations to it for a given patch. The solid black line in Fig.\ \ref{fig:hmc-cib-mkf} shows the MFs obtained from the average of these ten noise-contaminated CIB realisations. As the instrumental noise is subdominant at the beam resolution we chose in our analysis, we conclude that the MFs of the $25$ selected square patches closely follow the expected distribution from a Gaussian random field.
%%%%%%%%%%%%%%%%%%%%%%%%%%%%%%%%%%%%%%%%%%%%%%%%%%%%%%
%%%%%%%%%%%%%%%%%%%%%%%%%%%%%%%%%%%%%%%%%%%%%%%%%%%%%
%%%%%%%%%%%%%%%%%%%%%%%%%%%%%%%%%%%%%%%%%%%%%%%%%%%%%
\section{Scattering covariance statistics}\label{app:scat-coeff}

We used the SC statistics introduced in~\cite{cheng2024pnas} and~\cite{mousset2024aa}. The SC statistics correspond to the covariances of terms constructed from a combination of convolutions of the input random field $X$ with a set of pre-calculated wavelets and a non-linear modulus. The complex valued Morlet wavelet filters that we used, noted $\psi^{j,\gamma}$, are localised in pixel and harmonic space, and are characterised by a dyadic scale $j \in \left[0,\Jmax-1\right]$ and an orientation $\gamma \in \left [0,L-1\right]$. The wavelets probe characteristic scales of approximately $2^{j}$ pixels, and are oriented at an angle $\pi \gamma/L$ from the reference axis. $\Jmax < \log_2(N)$ is the number of dyadic scales available for a given image of $N \times N$ pixels. There are four types of SC statistics. The first $S_{1}$ and $S_2$ statistics jointly characterise the amplitude and sparsity of the process $X$ at a single oriented scale $\lambda = \paren*{j, \gamma}$. The higher-order wavelet moment estimators ($S_{3}$ and $S_{4}$) capture the interactions between two and three different oriented scales. 

Auto and cross-ST statistics can be computed between two random fields $X$ and $Y$. The cross SC statistics are defined as
\begin{equation}\label{eq:cross_scat_coeff}
\begin{aligned}
& S_{1}^{\times, \lambda_{1}}(X,Y) = \innerp*{\abs*{\paren*{X \ast \psi^{\lambda_{1}}}\overline{\paren*{Y \ast \psi^{\lambda_{1}}}}}}\,,\\
& S_{2}^{\times, \lambda_{1}}(X,Y) = \innerp*{\paren*{X \ast \psi^{\lambda_{1}}}\overline{\paren*{Y \ast \psi^{\lambda_{1}}}}}\,,\\
& S_{3}^{\times, \lambda_{1},\lambda_{2}}(X,Y) = \Cov*{X\ast\psi^{\lambda_{1}}, \abs*{Y\ast\psi^{\lambda_{2}}}\ast\psi^{\lambda_{1}}}\,,\\
& S_{3p}^{\times, \lambda_{1},\lambda_{2}}(X,Y) = \Cov*{Y\ast\psi^{\lambda_{1}}, \abs*{X\ast\psi^{\lambda_{2}}}\ast\psi^{\lambda_{1}}}\,,\\
 & S_{4}^{\times, \lambda_{1},\lambda_{2},\lambda_{3}}(X,Y) = \Cov*{\abs*{X\ast\psi^{\lambda_{3}}}\ast\psi^{\lambda_{1}}, \abs*{Y\ast\psi^{\lambda_{2}}}\ast\psi^{\lambda_{1}}}\,,
\end{aligned}
\end{equation}
where the asterisk stands for a convolution, the overbar is the complex conjugate, $\langle\cdot \rangle$ is the spatial average over the fields, and $\Cov{X,Y}$ is the estimated covariances of $X$ and $Y$. For auto-SC statistics, $S_1$ and $S_2$ are simplified to
\begin{equation}\label{eq:scat_coeff}
S_{1}^{\lambda_{1}}(X) = \innerp*{\abs*{X \ast \psi^{\lambda_{1}}}}   \ , \hspace{0.5cm}\mbox{and}\hspace{0.5cm}
S_{2}^{\lambda_{1}}(X) = \innerp*{\abs*{X \ast \psi^{\lambda_{1}}}^{2}}\,,
\end{equation}
while $S_3$ and $S_4$ were obtained by setting $X=Y$ in the previous equations, and the redundant $S_{3p}$ terms was removed. 
Following previous work~\citep{cheng2024pnas, mousset2024aa, campeti2025aa}, we normalised the $S_{3}^{\times}(X,Y)$ and $S_{4}^{\times}(X,Y)$ coefficients by their own $S_{2}^{\times}(X,Y)$ statistics,
\begin{equation}
\label{eq:normSC}
    \bar{S}_{3}^{\times, \lambda_{1},\lambda_{2}} = \frac{S_{3}^{\times, \lambda_{1},\lambda_{2}}}{\sqrt{S_{2}^{\times, \lambda_{1}} S_{2}^{\times,\lambda_{2}}}},\hspace{0.3cm}\mbox{and}\hspace{0.3cm}
    \bar{S}_{4}^{\times, \lambda_{1},\lambda_{2},\lambda_{3}} = \frac{S_{4}^{\times, \lambda_{1},\lambda_{2},\lambda_{3}}}{\sqrt{S_{2}^{\times, \lambda_{2}}  S_{2}^{\times,\lambda_{3}}}}.
\end{equation}

As we worked with fields with $256\times 256$ pixels, we set $\Jmax=5$ and $L=4$ for the wavelet filters using the kernel size of $5\times 5$ pixels. With these values of $\Jmax$ and $L$, we obtained $2522$ auto statistics and $2762$ cross statistics. At the end, the final summary statistics $\Phi$ that we used for the cross statistics were the concatenation of the mean of the fields $\innerp*{XY}$, its variance $\text{Var}\paren*{XY}$, and the normalised cross SC statistics, as
\begin{equation}\label{eq:cross-FinalPhi}
\begin{aligned}
    &\Phi\paren{XY} =\\
    &\lbrace \innerp*{XY}, \text{Var}\paren*{XY}, S_{1}^{\times, \lambda_{1}}, S_{2}^{\times, \lambda_{1}}, \bar{S}_{3}^{\times, \lambda_{1},\lambda_{2}}, \bar{S}_{3p}^{\times, \lambda_{1},\lambda_{2}}, \bar{S}_{4}^{\times, \lambda_{1},\lambda_{2},\lambda_{3}}  \rbrace.    
\end{aligned}
\end{equation}
For the auto statistics, the final summary statistics $\Phi$ were the concatenation of the mean of the field $\innerp*{X}$, its variance $\text{Var}\paren*{X}$, and the normalised SC statistics, yielding
\begin{equation}
\label{eq:auto-FinalPhi}
    \Phi\paren{X} = \lbrace \innerp*{X}, \text{Var}\paren*{X}, S_{1}^{\lambda_{1}}, S_{2}^{\lambda_{1}}, \bar{S}_{3}^{\lambda_{1},\lambda_{2}}, \bar{S}_{4}^{\lambda_{1},\lambda_{2},\lambda_{3}}  \rbrace.
\end{equation}
We computed these summary statistics using the Python package \foscat\footnote{\href{https://github.com/jmdelouis/FOSCAT}{https://github.com/jmdelouis/FOSCAT}}~\citep{delouis2022aa, campeti2025aa}. 
%%%%%%%%%%%%%%%%%%%%%%%%%%%%%%%%%%%%%%%%%%%%%%%%%%%%%
\section{Component-separation algorithm}\label{sec:wph-dust-cib}

In this section, we present the component-separation algorithm based on SC statistics to statistically separate the dust emission from the contamination at $353 \GHz$ \Planck observations. 
%%%%%%%%%%%%%%%%%%%%%%%%%%%%%%%%%%%%%%%%%%%%%%%%%%%%%
\subsection{Generative contamination maps}\label{sec:wph-cib}

We generated $300$ synthetic contamination maps ($\rsyn$) from the $25$ selected patches obtained in Sect.\ \ref{sec:cib-ps-all}. These synthetic maps were constructed using maximum entropy generative models conditioned by the $\Phi$ statistics, as defined in Eq.~\eqref{eq:auto-FinalPhi} (we refer to~\cite{bruna2018jmst, cheng2024pnas,mousset2024aa} for more details). 

We discuss below the steps that we followed to generate one of synthetic contamination maps. Starting from a white-noise realisation $u_{0}$, we performed a gradient descent in pixel space to minimise the following\footnote{In addition to the normalisation of the SC statistics described Eq.~\eqref{eq:normSC}, we take here the log of the $S_1$ and $S_2$ coefficients.} loss function:
\begin{equation}\label{eq:loss-part1}
\cL\paren*{u} = \loss{\Phi\paren*{\rB}-\Phi\paren*{u}}^{2},
\end{equation}
where $\loss{\,\cdot\,}$ is the Euclidean norm, and $\rB$ is one of the residual patch. The samples of the generative models are the maps $\rsyni$ obtained at the end of the optimisation, which verify
\begin{equation}
\Phi\paren*{\rB} \approx \Phi\paren*{\rsyni}.
\end{equation}

We optimised the loss function using the gradient descent algorithm in pixel space implemented in the \foscat package. The optimisation stopped when the difference of the loss functions in the two consecutive iterations was smaller than $10^{-6}$.
%%%%%%%%%%%%%%%%%%%%%%%%%%%%%%%%%%%%%%%%%%%%%%%%%%%%%
%%%%%%%%%%%%%%%%%%%%%%%%%%%%%%%%%%%%%%%%%%%%%%%%%%%%%%
\subsection{Component-separation algorithm}
\label{sec:wph-dust}

We started with the data map $m$ that we aimed to separate into the dust emission and the contamination. This was done by constructing maps that verified an ensemble of constraints that were constructed directly from the available data. The formalism for this statistical component-separation closely follows the formalisms presented in~\cite{delouis2022aa} and~\cite{auclair2024aa}. We used six constraints to recover a dust map that was statistically compatible with the information available in the \Planck data. The algorithm relies on a gradient descent on a running $u$ map, which at the end of the optimisation corresponds to the recovered $\ts$ map of the dust emission. We write below the constraints that this $\ts$ map should fulfill, before expressing these constraints as losses involving the running $u$ map.
The first constraint ensures that the recovered signal added to the contamination statistically matches the data. In terms of the summary statistics, the constraint is written as
\begin{equation}
\Phi\paren*{m} \simeq \innerp*{ \Phi\paren*{ \ts+ \rsyni} }_{i\in N},
\end{equation}
where $\langle\cdot \rangle_{i}$ is the ensemble average over the $N$ $\rsyn$ maps obtained in Sect.\ \ref{sec:wph-cib}.
The second constraint imposes that the cross-correlation between the data and the signal is conserved,
\begin{equation}
\Phi\paren*{m,\ts} \simeq \innerp*{ \Phi\paren*{ \ts+ \rsyni, \ts} }_{i\in N}.
\end{equation}
The third constraint imposes that the statistics of the residual map $\tr = \paren*{m - \ts}$ matches those of the $\rsyn$. This constraint ensures minimum leakage of the contamination into the recovered dust map. Mathematically,  
\begin{equation}
\Phi\paren*{m - \ts} \simeq \innerp*{\Phi\paren*{\rsyni}}_{i\in N}. 
\end{equation}
The fourth constraint enforces that the recovered dust map retains the same correlation with the total \NHI map as present in the data. The assumption here is that the contamination maps, being independent from the total \NHI, can be sampled again without modifying the estimate of the cross statistics. Mathematically, it is written as
\begin{equation}
\Phi\paren*{\NHI, m} \simeq \innerp*{\Phi\paren*{\NHI, \ts + \rsyni} }_{i\in N} \ .
\end{equation}
The fifth constraint is that the residual map is uncorrelated with the \NHI map, which yields
\begin{equation}
\Phi\paren*{\NHI, m-\ts} \simeq \innerp*{ \Phi\paren*{\NHI, \rsyni} }_{i\in N}.
\end{equation}
Finally, the sixth constraint imposes that the residual map is uncorrelated with the dust map separately, 
\begin{equation}
\Phi\paren*{\ts, m-\ts} \simeq \innerp*{ \Phi\paren*{\ts, \rsyni} }_{i\in N} \ .
\end{equation}
The last four constraints enforce minimum leakage of the dust signal into the contamination map at the angular scales at which the amplitude of the dust signal is comparable to that of the contamination. 

Following \cite{delouis2022aa}, the mathematical loss functions corresponding to these constraints are respectively written as
% %%%%%%%%%%%%%%%%%%%%%%%%%%%%%%%%%%%%%%%%%%%%%%%%%%%%%%
\begin{equation}\label{eq:loss-part2}
\begin{aligned}
 \cL_{1}\paren*{u} = &\loss*{\frac{\Phi(m) - \Phi(u) - B_{1}}{\sigma_{\Phi(u+\rsyni)}}}^{2}, \\
 & B_{1}= \innerp*{\Phi(u+\rsyni)- \Phi(u)}_{i\in N}\\
\cL_{2}\paren*{u} = &\loss*{\frac{\Phi(m,u) - \Phi(u,u) - B_{2}}{\sigma_{\Phi(u+\rsyni, u)}}}^{2},\\
&  B_{2}= \langle\Phi(u+\rsyni, u)- \Phi(u, u)\rangle_{i\in N},\\
\cL_{3}\paren*{u} = &\loss*{\frac{\Phi(m-u) - \innerp*{\Phi(\rsyni)}_{i\in N}}{\sigma_{\Phi(\rsyni)}}}^{2}, \\
\cL_{4}\paren*{u} = &\loss*{\frac{\Phi(\NHI, m) - \Phi(\NHI, u) - B_{4}}{\sigma_{\Phi(\NHI, u+ \rsyni)}}}^{2},\\
& B_{4}= \innerp*{\Phi(\NHI, u+\rsyni)- \Phi(\NHI, u)}_{i\in N},\\
\cL_{5}\paren*{u} = &\loss*{\frac{\Phi(\NHI, m-u) - \langle\Phi(\NHI, \rsyni)\rangle_{i\in N}}{\sigma_{\Phi(\NHI, \rsyni)}}}^{2}, \\
\cL_{6}\paren*{u} = &\loss*{\frac{\Phi(u, m-u) - \langle\Phi(u, \rsyni)\rangle_{i\in N}}{\sigma_{\Phi(u, \rsyni)}}}^{2}.
\end{aligned}
\end{equation}
%%%%%%%%%%%%%%%%%%%%%%%%%%%%%%%%%%%%%%%%%%%%%%%%%%%%%%
We minimised the total loss function, given as
\begin{equation}\label{eq:loss-total2}
\cL\paren*{u} = \cL_{1}\paren*{u} + \cL_{2}\paren*{u} + \cL_{3}\paren*{u}+ \cL_{4}\paren*{u} + \cL_{5}\paren*{u}+ \cL_{6}\paren*{u}.
\end{equation}

%%%%%%%%%%%%%%%%%%%%%%%%%%%%%%%%%%%%%%%%%%%%%%%%%%%%%%
For the detailed concept behind the bias $B$ and standard deviation $\sigma$, we refer to~\cite{delouis2022aa}. For example, the loss $\cL_{1}$ is computed as the normalised chi-square distribution between $\Phi (m)$ and $\innerp*{\Phi (u + \rsyni)}_{i\in N}$. However, for computational efficiency, it relies on an bias $B_1$, which is estimated only after a certain number of iterations. This avoids the need to trace $\innerp*{\Phi (u + \rsyni)}_{i\in N}$ throughout the optimisation. This bias is only necessary when the loss involves an ensemble average over the $\rsyni$ maps.

The different initial condition maps for the gradient descent were made from the $m$ map smoothed with a $W\arcm$ (FWHM) Gaussian beam, where $W$ was chosen randomly from a uniform distribution $\mathcal{U}[100, 200]$. Starting from the initial map $u_{0}$, we minimised the total loss function using the gradient descent algorithm in \foscat. The final dust map $\ts$ corresponded to $u$ at the end of the optimisation, and the residual map was obtained as $\tr =\paren*{m-\ts}$. In practice, we updated the $B_1$, $B_2$ and $B_4$ biases and all variance terms every 150 iterations. We ran this optimisation until the difference between two consecutive losses in an epoch was $\Delta \cL < 0.002$. The typical number of epochs required to reach the $\Delta \cL$ varied from five to eight, depending on the data map. Numerically, we found that decreasing $\Delta \cL$ beyond $0.002$ did not improve the recovered dust map. The total computation time ranged from 15 to 20 minutes on a single-node GPU cluster (\texttt{NVIDIA A30}), depending on the number of iterations performed.
%%%%%%%%%%%%%%%%%%%%%%%%%%%%%%%%%%%%%%%%%%%%%%%%%%%%%
\section{\Planck simulations}\label{sec:planck-simulation}

We first validated the component-separation algorithm described in Sect.\ \ref{sec:wph-dust} on simulated \Planck maps on a 2D square patch.
%%%%%%%%%%%%%%%%%%%%%%%%%%%%%%%%%%%%%%%%%%%%%%%%%%%%%
\subsection{Synthetic contamination maps}\label{sec:syn-wph-cib}

\begin{figure}
\centering
\includegraphics[width=\columnwidth]{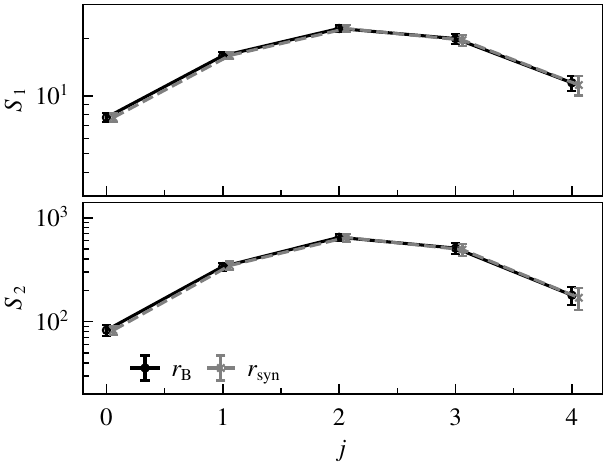}
\caption{Comparison of the $S_{1}$ and $S_{2}$ statistics of mean and standard deviation of $25$ contamination maps obtained from the template-fit approach $\rB$ regions (black circles) and 300 synthetic contamination maps $\rsyn$ (grey crosses). The grey crosses are shifted in the $x$-axis. }
\label{fig:hmc-wph-sc}
\end{figure}

We synthesised $12$ statistically identical realisations from each sky patch considered in Sect.~\ref{sec:hmc-cib}, totaling $N=300$ realisations of $\rsyn$ maps using the summary statistics, as discussed in Sect.\ \ref{sec:wph-cib}. The synthetic contamination maps retained the all the statistical properties of the $\rB$ and had a mean standard deviation $\sigma_{\rsyn}$ of $9.0\,\kJysr$, and the estimated $1\sigma$ variation in the $\sigma_{\rsyn}$ was $0.4\,\kJysr$. These synthetic maps followed the Gaussian distribution and had the same one-dimensional probability distribution function as the $\rB$ maps. We also computed the angular power spectra and the three MFs. They followed the mean $\rB$ statistics. For brevity, we only show the mean and standard deviation of the $S_{1}$ and $S_{2}$ coefficients obtained from the $25\,\rB$ and $300\,\rsyn$ maps in Fig.\ \ref{fig:hmc-wph-sc}. We show the mean and standard deviation of the other two normalised coefficients $\bar{S}_{3}$ and $\bar{S}_{4}$ in Fig.\ \ref{fig:hmc-wph-sc-S3}. We used these $300\,\rsyn$ maps to compute the variance on its summary statistics.
%%%%%%%%%%%%%%%%%%%%%%%%%%%%%%%%%%%%%%%%%%%%%%%%%%%%%
%%%%%%%%%%%%%%%%%%%%%%%%%%%%%%%%%%%%%%%%%%%%%%%%%%%%%
%%%%%%%%%%%%%%%%%%%%%%%%%%%%%%%%%%%%%%%%%%%%%%%%%%%%%
%%%%%%%%%%%%%%%%%%%%%%%%%%%%%%%%%%%%%%%%%%%%%%%%%%%%%
\subsection{Simulated map at $353 \GHz$}

We chose a test patch centred at Galactic coordinates ${(l, b) = (45.0\degree, -54.3\degree)}$ of sky area $222 \deg^2$. The mean \HI column density over the entire patch was $\innerp{\NHI}=3.3 \times \hiunit$. We extracted the same sky patch from the CSFD map and scaled it with a constant factor to produce an input dust map. We kept the contamination map fixed and varied the dust amplitude to produce different $\mathrm{S/N}$ maps. The constant scaling factor controls the $\mathrm{S/N}$s of the input dust map with respect to the contamination.

We validated our algorithm for $\mathrm{S/N}$s 3 to 9. We discuss the $\mathrm{S/N}=3$ results in detail and highlight the primary differences among the others. We call the scaled 2D input dust map $\ssim$, to which we added a realisation of the contamination map ($\rsim$) from the generative model $\rsyn$ obtained in Sect.\ \ref{sec:syn-wph-cib} to produce the contaminated map ($\msim = \ssim + \rsim$). The Pearson correlation coefficient $\rho$ of $\ssim$ with the \NHI map for $\mathrm{S/N}=3$ is $0.94$ and was reduced to $0.88$ after we added the contamination map $\rsim$. The top panel of Fig.~\ref{fig:sim-wph-patch} shows the input dust map at $\mathrm{S/N}=3$, one random realisation of the contamination, and the total contaminated simulated map. $\msim$ map is patchy owing to contamination. 
%%%%%%%%%%%%%%%%%%%%%%%%%%%%%%%%%%%%%%%%%%%%%%%%%%%%%%
%%%%%%%%%%%%%%%%%%%%%%%%%%%%%%%%%%%%%%%%%%%%%%%%%%%%%
\subsection{Validating the component-separation algorithm}

\begin{figure}[!h]
\centering
\includegraphics[width=\columnwidth]{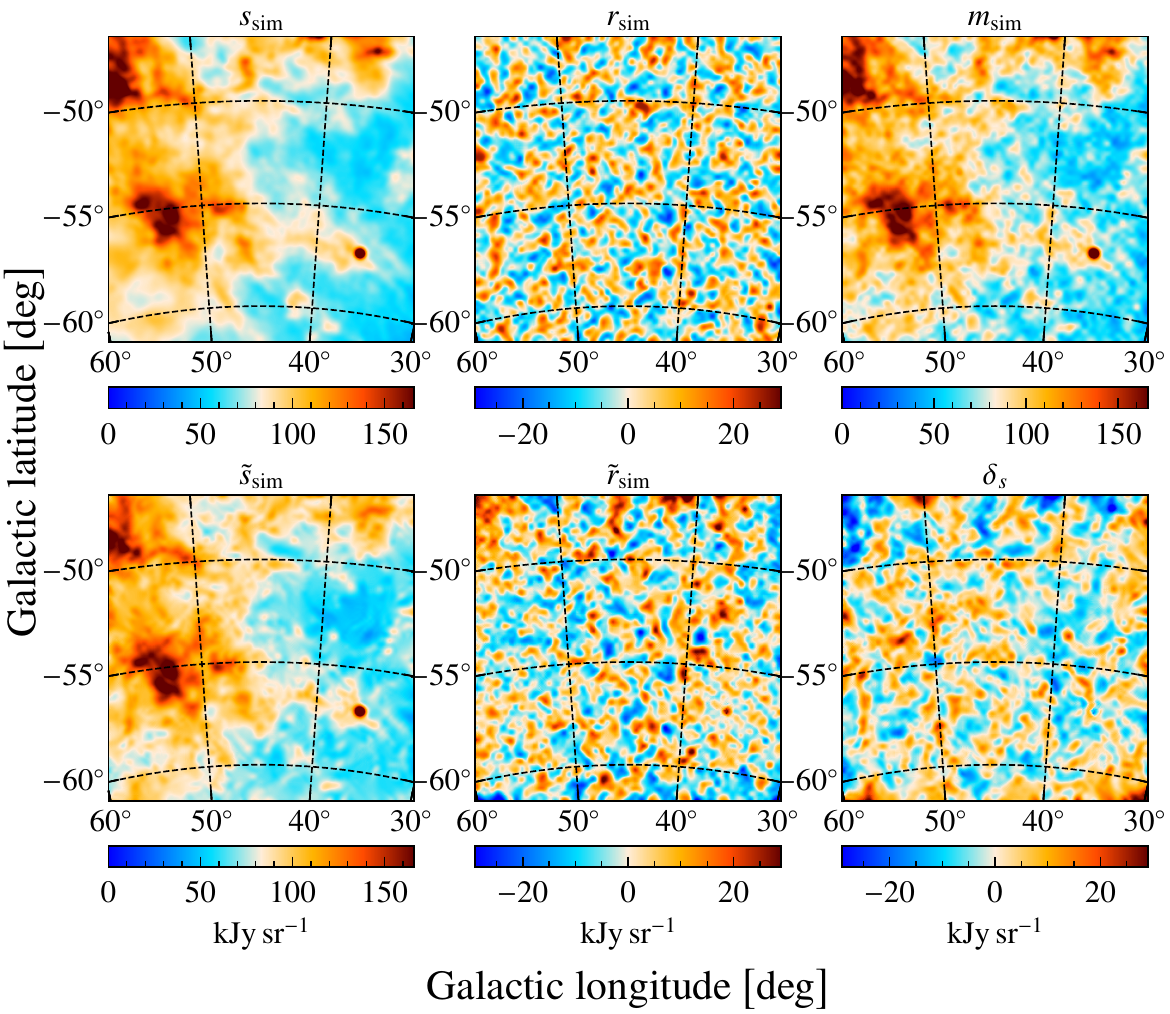}
\caption{\textit{Top}: Input maps at $\mathrm{S/N}=3$ centred around the sky patch at $(l, b)= (45.0\degree, -54.3\degree)$. The columns from \textit{left to right} show the input dust map ($\ssim$), the input contamination map ($\rsim$), and the total simulated map ($\msim$).   \textit{Bottom}: Columns from left to right: Component-separated dust map ($\tssim$), residual map ($\trsim$), and difference between input and recovered dust map ($\delta_s$). 
} 
\label{fig:sim-wph-patch}
\end{figure}

\begin{figure}[!h]
\centering
\includegraphics[width=\columnwidth]{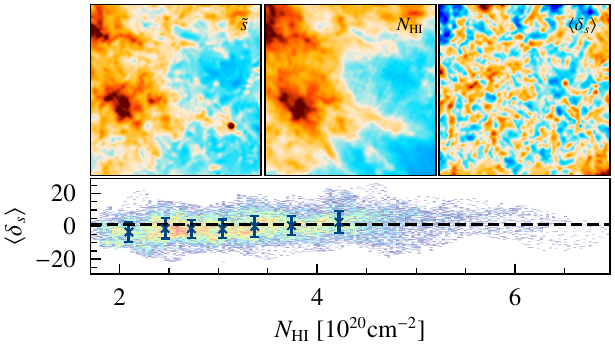}
\caption{\textit{Top:} $\tssim$ (left), \HI column density map (middle), and the mean of $25$ realisations of $\delta_{s}$ (right). \textit{Bottom:} 2D correlation between \NHI and $\langle \delta_{s} \rangle$. The blue crosses and the error bars show the median values and standard deviations of mean $\langle \delta_{s} \rangle$ in the ordered bins of \NHI.} 
\label{fig:sim-wph-diff-patch}
\end{figure}

%%%%%%%%%%%%%%%%%%%%%%%%%%%%%%%%%%%%%%%%%%%%%%%%%%%%%
We applied the component-separation algorithm described in Sect.\ \ref{sec:wph-dust} to the contaminated map to extract a statistical realisation of the dust signal. By performing the component-separation under different initial conditions, we obtained $25$ recovered dust maps. The bottom panel of Fig.~\ref{fig:sim-wph-patch} shows one of the recovered dust maps ($\tssim$), the residual map ($\trsim = \msim - \tssim$), and the difference between the input and recovered dust ($\delta_s = \tssim - \ssim$). We clearly see some Galactic residuals near the boundaries of $\trsim$ map. To avoid these boundary pixels, we restricted our analysis to the unmasked pixels within the binary mask (shown in the left panel of Fig.\ \ref{fig:square-mask}) to compute the Pearson correlation coefficients.  The value of $\rho$ between $\ssim$ and $\tssim$ was found to be $0.97$, showing a tight pixel-to-pixel correlation between the input and recovered dust maps. Visually, the small-scale features dominated by contamination are well separated from the large-scale dust emission. The $\delta_{s}$ map has no visible large-scale features at the map level. The correlation coefficient $\rho=0.91$ between $\tssim$ and \NHI is the same as was obtained between $\ssim$ and \NHI. This shows that the component-separation algorithm removes most of the contamination from the total map. 
%%%%%%%%%%%%%%%%%%%%%%%%%%%%%%%%%%%%%%%%%%%%%%%%%%%%%%
%%%%%%%%%%%%%%%%%%%%%%%%%%%%%%%%%%%%%%%%%%%%%%%%%%%%%%
%%%%%%%%%%%%%%%%%%%%%%%%%%%%%%%%%%%%%%%%%%%%%%%%%%%%%
\begin{figure} [!htbp]
   \centering
   \includegraphics[width=\columnwidth, keepaspectratio=True]{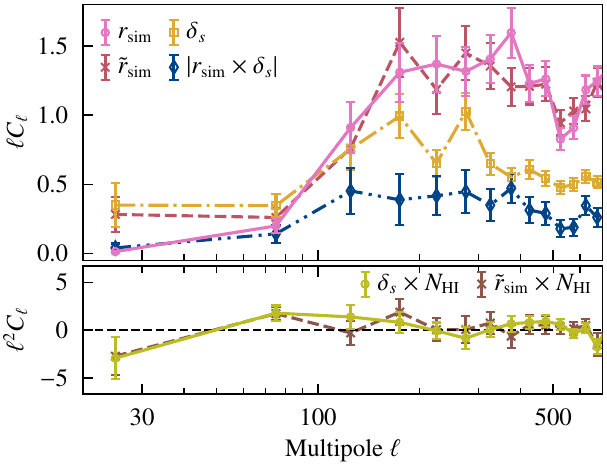}
      \caption{\textit{Top}: $\ell\Cl$ spectra for $\rsim$ (solid pink), $\trsim$ (dashed~red), $\delta_{s}$ (dashed-dotted orange), and magnitude of the cross-spectrum for $\rsim$ and $\delta_{s}$ (dashed-dot-dotted blue) in \kJysrsq.
      \textit{Bottom}: Cross-spectra ($\ell^{2}\Cl$) of $\delta_s$ (solid green) and $\trsim$ (dashed brown) with the \NHI map, in $\kJysr\paren*{\hiunit}$.}\label{fig:sim-wph-cib-ps}
\end{figure}   
%%%%%%%%%%%%%%%%%%%%%%%%%%%%%%%%%%%%%%%%%%%%%%%%%%%%%%%
The top panel of Fig.\ \ref{fig:sim-wph-cib-ps} shows the auto spectra $\ell \Cl$ of $\rsim$, $\trsim$, and $\delta_{s}$, along with the magnitude of the cross-spectrum of $\rsim $ with $\delta_{s}$. The bottom panel presents the cross-spectra $\ell^{2}\Cl$ of $\delta_s$ and $\trsim$ with the \NHI map. We applied the apodised mask (Fig.\ \ref{fig:square-mask}, right) to account for non-periodic boundaries. The power spectrum of the $\rsim$ follows an $\Cl \propto \ell^{-1.3}$ spectrum, which is consistent with the CIB model of \cite{planck-XXX:2014}. As our component-separation algorithm is a statistical method, the $\delta_s$ map captures the phase mismatch between the input and the output dust maps. The $\delta_s$ map is not correlated with \NHI map, as is captured by the cross-power spectrum of $\delta_s$ and \NHI. The average $\delta_s$ map obtained from $25$ realisations is shown in Fig.\ \ref{fig:sim-wph-diff-patch} (third column, top panel). The average $\delta_s$ map is also statistically uncorrelated with \NHI at the pixel level (bottom panel of Fig.\ \ref{fig:sim-wph-diff-patch}). We show the comparison between the input and output dust spectra in Fig.\ \ref{fig:sim-wph-dust-ps}.
%%%%%%%%%%%%%%%%%%%%%%%%%%%%%%%%%%%%%%%%%%%%%%%%%%%%%%
%%%%%%%%%%%%%%%%%%%%%%%%%%%%%%%%%%%%%%%%%%%%%%%%%%%%%%
%%%%%%%%%%%%%%%%%%%%%%%%%%%%%%%%%%%%%%%%%%%%%%%%%%%%%%
%%%%%%%%%%%%%%%%%%%%%%%%%%%%%%%%%%%%%%%%%%%%%%%%%%%%%
\begin{figure} [!hb]
\centering
\includegraphics[width=\columnwidth]{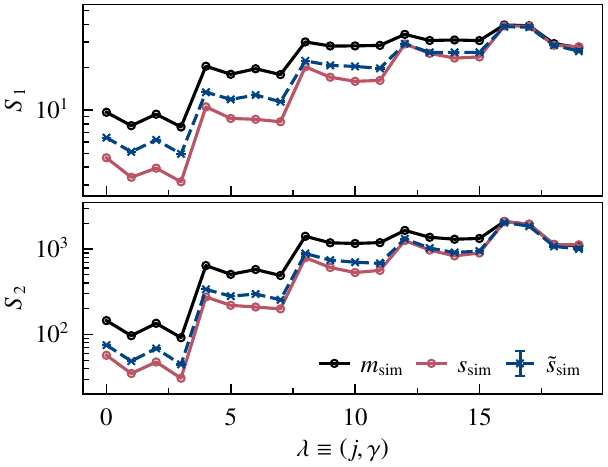}
\caption{Comparison of $S_{1}$ and $S_{2}$ statistics of $\msim$ (black circles), $\ssim$ (red circles), and the $\tssim$ (blue crosses with dashed line).}
\label{fig:sim-wph-sc}
\end{figure}   
%%%%%%%%%%%%%%%%%%%%%%%%%%%%%%%%%%%%%%%%%%%%%%%%%%%%%
%%%%%%%%%%%%%%%%%%%%%%%%%%%%%%%%%%%%%%%%%%%%%%%%%%%%%
\begin{figure} [!htbp]
\centering
\includegraphics[width=\columnwidth]{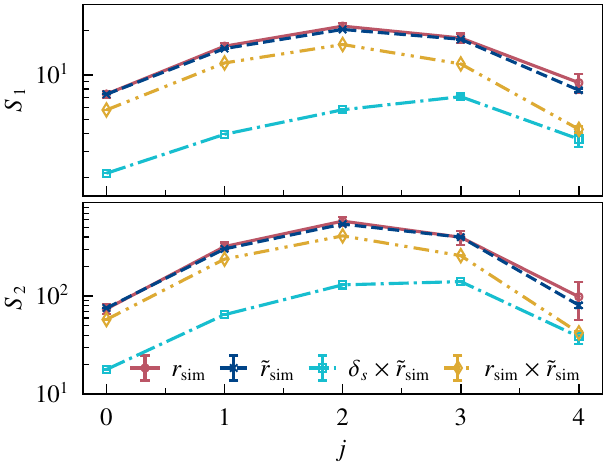}
\caption{Comparison of the angle-averaged $S_{1}$ and $S_{2}$ statistics of $\rsim$ (red circles) and $\trsim$ (blue crosses with the dashed line). The cyan squares with the dashed-dotted line show the cross coefficients between $\delta_s$ and $\trsim$, and the orange diamonds with the dashed-dot-dotted line show that between $\rsim$ and $\trsim$.}
\label{fig:sim-sc-cib}
\end{figure}   
%%%%%%%%%%%%%%%%%%%%%%%%%%%%%%%%%%%%%%%%%%%%%%%%%%%%%
%%%%%%%%%%%%%%%%%%%%%%%%%%%%%%%%%%%%%%%%%%%%%%%%%%%%%%
\begin{figure} [!htbp]
\centering
\includegraphics[width=\columnwidth, keepaspectratio=True]{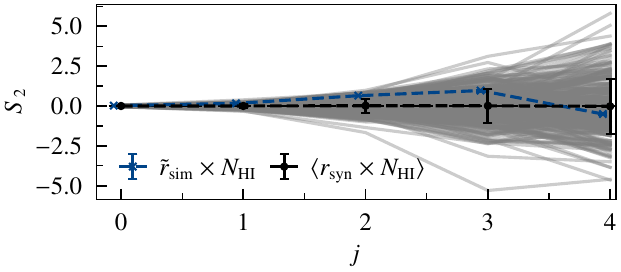}
\caption{Comparison of the angle-averaged $S_{2}$ cross statistics with the $\NHI$ map: $\trsim$ (dashed blue) and the mean over $300\,\rsyn$ (black). The recovered contamination is consistent with the expected statistics within $1\sigma$. The blue crosses are slightly shifted along the $x$-axis. The grey lines show all $300\,\rsyn$ realisations.}
\label{fig:sim-wph-rsim-hi}
\end{figure}   
%%%%%%%%%%%%%%%%%%%%%%%%%%%%%%%%%%%%%%%%%%%%%%%%%%%%%%

Next, we calculated the SC statistics ($S_1$, $S_2$, $\bar{S}_{3}$, and $\bar{S}_{4}$) of these maps. The mean and standard deviation of $S_{1}$ and $S_{2}$ for five different $j$ values and four different $\gamma$ values, computed from $25$ recovered dust maps, are shown in Fig.\ \ref{fig:sim-wph-sc}. The $x$-axis represents $\lambda\equiv \paren*{j, \gamma}$, where $\lambda=0$ corresponds to the smallest scale ($j=0$) and the first wavelet orientation. Our component-separation algorithm accurately recovers these statistical properties of the dust map as a function of wavelet scales and orientations. The contamination in $\msim$ leads to higher $S_{1}$ and $S_{2}$ at small scales (or low $j$ values) compared to $\ssim$. The algorithm recovers almost all the scales of $\ssim$, except for the smallest scales at $\mathrm{S/N}=3$. We show the remaining two SC statistics $\bar{S}_{3}$ and $\bar{S}_{4}$ in Fig. \ \ref{fig:sim-wph-sc-S3}.  At $\mathrm{S/N} \geqslant 5$, the algorithm recovers $S_1$ and $S_2$ statistics for large and small scales. In Fig.\ \ref{fig:sim-wph-sc-S3-snr5} we present the SC statistics for $\mathrm{S/N}=5$. The smallest scale ($j=0$) between the input and output for all statistics agrees well. 
Figure \ref{fig:sim-sc-cib} shows the angle-averaged $S_{1}$ and $S_{2}$ statistics of $\rsim$ and $\trsim$ for different values of $j$. As the contamination maps is statistically isotropic, there is no preferred orientation at which we expect to see more power. For $\rsim$, we computed the $1\sigma$ standard deviation from $300\,\rsyn$ maps. In the same plot, we also show the same angle-averaged $S_{1}$ and $S_{2}$ cross statistics of the $\delta_s$ and $\trsim$ maps. The cross $S_{1}$ and $S_{2}$ statistics between $\rsim$ and $\trsim$ broadly follow the same statistics as $\rsim$. This means that the $\trsim$ map retains the phase information of the input $\rsim$ up to a certain percentage. Even though $\delta_s$ visually appears to be similar to $\rsim$, the $S_2$ amplitude of $\delta_s \times \trsim$ is much smaller than the corresponding amplitude of $\rsim \times \rsim$ (or $\trsim \times \trsim$) at small scales corresponding to low $j$ values. The $1\sigma$ deviation on the auto and cross statistics with the recovered maps was computed from the $25$ realisations of $\tr$. Figure~\ref{fig:sim-wph-rsim-hi} shows the angle-averaged cross statistics $S_{2}$ between $\trsim$ and \NHI map. The error bar on $S_2$ was computed from $25$ realisations of the $\trsim$, starting from different initial conditions. For completeness, we computed the expected cross statistics $S_{2}$ between $\rsyn$ and \NHI map. The mean value of the $S_2$ statistics computed from $300\,\rsyn$ maps is consistent with zero, and the $1\sigma$ error bars capture patch-to-patch variations of the same statistics. Since we added a single realisation of the contamination to the input dust map, the error bar on the $\trsim \times \NHI$ cross statistics $S_{2}$ only includes the statistical noise from the component-separation algorithm. 

%%%%%%%%%%%%%%%%%%%%%%%%%%%%%%%%%%%%%%%%%%%%%%%%%%%%%%
\begin{figure} [!htbp]
\centering
\includegraphics[width=\columnwidth]{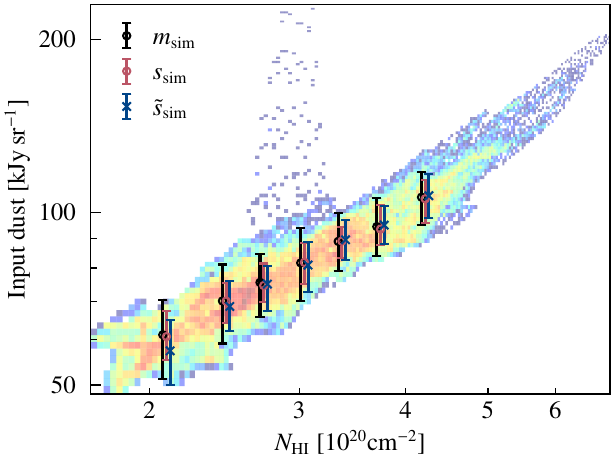}
\caption{2D histogram showing the joint distribution of $\ssim$ and $\NHI$. The black circles correspond to the median values, and the upper and lower limits correspond to the 16 and 84 percentile of $\msim$ of the data in each ordered bin of $\NHI$. Similarly, the red circles show $\ssim$, and the blue crosses show $\tssim$.} 
\label{fig:sim-wph-hi-corr}
\end{figure}   
%%%%%%%%%%%%%%%%%%%%%%%%%%%%%%%%%%%%%%%%%%%%%%%%%%%%%
%%%%%%%%%%%%%%%%%%%%%%%%%%%%%%%%%%%%%%%%%%%%%%%%%%%%%%
In Fig.\ \ref{fig:sim-wph-hi-corr} we plot the 2D histogram highlighting the joint distribution of $\ssim$ and $\NHI$, considering only the pixels those falls within the binary mask. This shows that the $\ssim$ has a significant amount of scatter with $\NHI$ over the entire patch, and the correlation between them becomes non-linear above $\NHI > 4 \times 10^{20} \cm^{-2}$. Next, we binned the data into seven bins in order of increasing values of \NHI and computed the median values of $\msim$, $\ssim$, and $\tssim$ over each bin. The upper and lower limit of the error bars at each bin correspond to the $16$ and $84$ percentile of the three quantities $\msim$, $\ssim$, and $\tssim$. 
This plot shows that the distributions are very similar for these different maps, although there appears to be a consistent reduction in the standard deviation of the distribution at fixed \NHI values from $\msim$ to $\tssim$. This effect is more pronounced in low \NHI regions, in which the contamination is comparable to the dust signal. The reduction of the error bars from $\msim$ to $\tssim$ is consistent with the expected outcome of removing contaminants.
%%%%%%%%%%%%%%%%%%%%%%%%%%%%%%%%%%%%%%%%%%%%%%%%%%%%%%
%%%%%%%%%%%%%%%%%%%%%%%%%%%%%%%%%%%%%%%%%%%%%%%%%%%%%%

Next, we varied the S/N of the dust signal with respect to contamination. For each S/N, we quantified the reliability of the component-separation algorithm. For $\mathrm{S/N} \geqslant 3$, we demonstrate that our component-separation algorithm successfully and efficiently separates the dust signal from the contamination at the level of summary statistics.
The algorithm only recovers the dust signal at large scales for low S/Ns ($\mathrm{S/N} \simeq 1$), leaving excess power at small scales in the recovered dust map. We applied the same procedure to other sky regions for different S/Ns and reached the same conclusions. 
%%%%%%%%%%%%%%%%%%%%%%%%%%%%%%%%%%%%%%%%%%%%%%%%%%%%%%
%%%%%%%%%%%%%%%%%%%%%%%%%%%%%%%%%%%%%%%%%%%%%%%%%%%%%
\section{Application to \Planck data}\label{sec:results}
%%%%%%%%%%%%%%%%%%%%%%%%%%%%%%%%%%%%%%%%%%%%%%%%%%%%%%
\subsection{Component-separated maps and its summary statistics}\label{sec:data_results_maps}

%%%%%%%%%%%%%%%%%%%%%%%%%%%%%%%%%%%%%%%%%%%%%%%%%%%%%%
%%%%%%%%%%%%%%%%%%%%%%%%%%%%%%%%%%%%%%%%%%%%%%%%%%%%%%
\begin{figure*}
\begin{center}
    \includegraphics[width=0.8\linewidth, keepaspectratio=True]{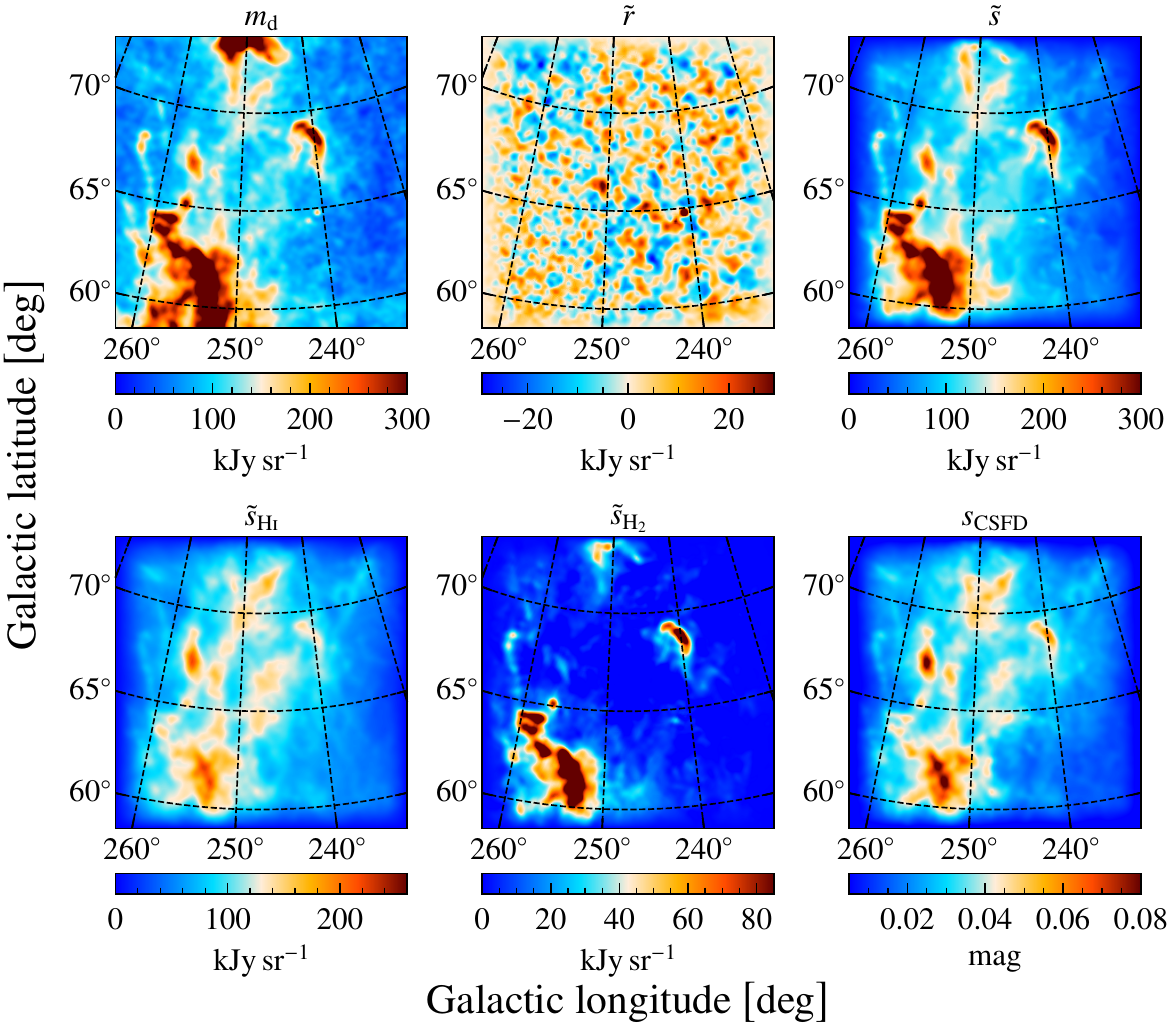}
\caption{\textit{Top:} \Planck $353 \GHz$ map (CMB and global offset subtracted; left), $\tr$ (middle) and  $\ts$ (right) after component-separation in the same sky region as in Fig.\ \ref{fig:sim-wph-patch}. \textit{Bottom:} Decomposed $\ts_{\HI}$ (left) and $\ts_{\molH}$ maps (middle). The $\scfd$ map at $100\,\mum$ (in $\rm{mag}$ units) in the same sky region is shown in the right panel.}\label{fig:result-wph-data}
\end{center}
\end{figure*}     
%%%%%%%%%%%%%%%%%%%%%%%%%%%%%%%%%%%%%%%%%%%%%%%%%%%%%%
%%%%%%%%%%%%%%%%%%%%%%%%%%%%%%%%%%%%%%%%%%%%%%%%%%%%%%
In this section, we primarily focus on the component-separation results on a sky patch of 222 deg$^2$ area centred at Galactic coordinates $(l, b) = (247.5\degree, 66.4\degree)$ for brevity. We first subtracted the global offset term of $119.3\,\kJysr$ as obtained from the template-fit approach (see Sect.\ \ref{sec:hmc-cib}) from the CMB-subtracted \Planck data to focus on the statistical separation of the dust emission and CIB contamination using the SC-based statistics. We refer to this map as $\md$. The mean \HI column density measured across the entire map is $\innerp{\NHI}=2.3 \times \hiunit$. The Pearson correlation coefficient $\rho$ between the $\md$ and $\NHI$ maps is $0.84$.

We applied the component-separation algorithm to extract a statistical realisation of the dust signal. We followed the same steps as described in Sect.\ \ref{sec:planck-simulation} and obtained $25$ different realisations of dust map. The top panel of Fig.\ \ref{fig:result-wph-data} shows the input map ($\md$), recovered residual map ($\tr$) and a single realisation of the dust map ($\ts$). The recovered dust map retains most of the large-scale features present in the \Planck data and filters out the small-scale fluctuations from the contamination. The mean dust emissivity over the entire sky patch was computed as the mean of the ratio of the recovered dust over \NHI, defined as $\langle \epsilon \rangle = \langle \ts/\NHI \rangle$. The mean value of the dust emissivity is $45\,\kJysr\paren*{\hiunit}$, which is consistent with the value obtained by~\citet{planck-XXIV:2011, planck-XVII:2014}, and~\citet{adak2024mnras}. 

\begin{figure} [!htbp]
\centering
\includegraphics[width=\columnwidth]{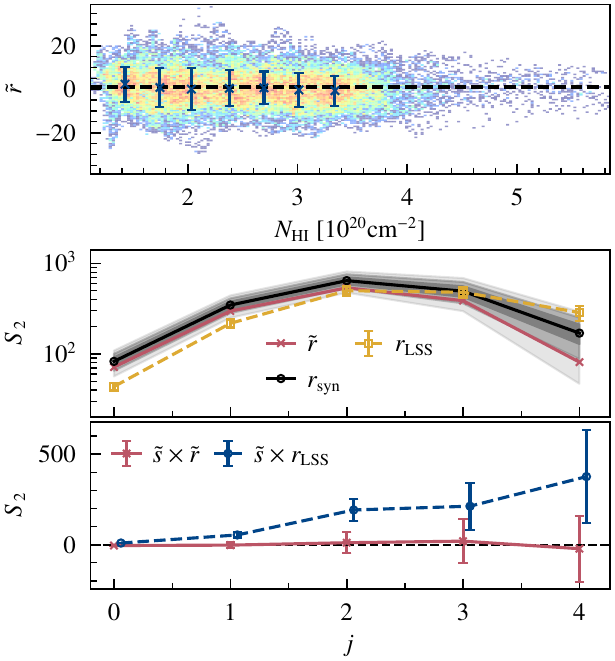}
\caption{\textit{Top:} Correlation plot of $\tr$ with \NHI map. The blue data points and the error bars show the zero median and standard deviation of $\tr$ in the ordered bins of \NHI. The standard deviations are consistent in all the \NHI bins. \textit{Middle:} Comparison of angle-averaged $S_2$ statistics of $\tr$ (red cross), $\rlss$ (square dashed orange line) and the ensemble average of $300\,\rsyn$ maps (black circles) along with $1\sigma$, $2\sigma$, and $3\sigma$ error bands. \textit{Bottom:} Red crosses with the solid line show cross $S_{2}$ coefficients between $\ts$ and $\tr$, and the blue circles with the dashed line show the cross $S_{2}$ coefficients between $\ts$ and $\rlss$. The blue circles are slightly shifted in the $x$-axis.} 
\label{fig:result-wph-cib}
\end{figure}

Next, we present the statistics of the residual map ($\tr$). The top panel in Fig.\ \ref{fig:result-wph-cib} shows the joint distribution of $\tr$ and the \NHI map. The blue data points are the median values, and the error bars correspond to $1\sigma$ standard deviation in seven bins ordered in increasing values of \NHI. The median values are consistent with zero, showing no significant leakage of $\ts$ into the residual map. The plot also confirms that our SC statistics-based component-separation algorithm works well for the $\mathrm{S/N}\approx 8.4$ region. The standard deviation of $\sigma_{\tr}$ over the masked region is $8.1\,\kJysr$, which is consistent with the expected value from $300\,\rsyn$ maps.
%%%%%%%%%%%%%%%%%%%%%%%%%%%%%%%%%%%%%%%%%%%%%%%%%%%%%%
%%%%%%%%%%%%%%%%%%%%%%%%%%%%%%%%%%%%%%%%%%%%%%%%%%%%%%
In the middle panel of Fig.\ \ref{fig:result-wph-cib}, we show the $S_2$ comparison of the $\tr$ map in the \Planck field and the expected distribution from $300\,\rsyn$ maps. The amplitudes of the $S_2$ statistics match well at low $j$ values, except at the very large scale ($j=4$).  The difference in $S_2$ statistics between the $\tr$ and $300\,\rsyn$ map is at the level of $2.3\sigma$ for $j=4$. We also compared the $\rlss$ map  at $100\,\mum$ from~\cite{chiang2023apj} at that field. We multiplied the $\rlss$ map (in $\kJysr$ units) by a factor of $0.3$ and computed the $S_2$ coefficient. We found that small-scale powers are missing from the $\rlss$ map, but the intermediate and large scales are consistent within the error bands from the synthetic maps. We correlated the $\ts$ map with $\tr$ map and found a cross-correlation coefficient $\rho=-0.06$ between the two maps. We show the cross $S_{2}$ statistics of $\ts$ and $\tr$ in the bottom panel of Fig.\ \ref{fig:result-wph-cib} to ensure that the recovered maps are statistically uncorrelated at the level of $S_{2}$ statistics. The $1\sigma$ deviations on the coefficients were computed from the cross $S_{2}$ statistics of $\ts$ and the $300\,\rsyn$ maps. We also computed the cross statistics $S_2$ between the $\ts$ and $\rlss$ map for the same sky patch. The $1\sigma$ deviations on $S_2$ were computed from the cross statistics of $\ts$ with 94 independent sky patches with a size of $14.9\degree\times 14.9\degree$ of the \cite{chiang2023apj} CIB/LSS map. All the scales are consistent with the zero within the $1\sigma$ error bar, except for the $j=2$ scale, where the deviation is at the level of $3.2\sigma$. The cross-correlation coefficient of the $\ts$ map with $\rlss$ map is $\rho=0.03$. 
%%%%%%%%%%%%%%%%%%%%%%%%%%%%%%%%%%%%%%%%%%%%%%%%%%%%%%
%%%%%%%%%%%%%%%%%%%%%%%%%%%%%%%%%%%%%%%%%%%%%%%%%%%%%%

We also performed a non-Gaussianity test to check for any significant leakage of the Galactic residuals into the $\tr$ map. The results of the non-Gaussianity test are presented in Appendix \ref{app:data-extra-temp}. Additionally, we compared the statistics of the contamination maps from two different component-separation algorithms: the template-fit approach, and the SC-based statistics. In Appendix \ref{app:data-results-comp} we present the component-separation results for the same \Planck field as used to validate the component-separation algorithm in Sect.\ \ref{sec:planck-simulation}.
%%%%%%%%%%%%%%%%%%%%%%%%%%%%%%%%%%%%%%%%%%%%%%%%%%%%%%
%%%%%%%%%%%%%%%%%%%%%%%%%%%%%%%%%%%%%%%%%%%%%%%%%%%%
\subsection{Separation in \HI and \molH emission}\label{sec:data-wph-nhi-nh2}

To understand the morphology of the recovered dust emission, we decomposed it into two components: dust associated with \NHI ($\ts_{\HI}$), and dust associated with \NmolH ($\ts_{\molH}$),
\begin{equation}
    \ts = \ts_{\HI} + \ts_{\molH}, \label{eq:data-nhi-nh2}
\end{equation}
where $\ts_{\HI} = \epsilon_{\HI,353} \NHI$, and $\epsilon_{\HI,353}$ is the dust emissivity of the \HI-correlated dust emission at $353 \GHz$. In our case, $\epsilon_{\HI,353}$ is a constant value over the selected sky patch.
%%%%%%%%%%%%%%%%%%%%%%%%%%%%%%%%%%%%%%%%%%%%%%%%%%%%%%
 We estimated the mean $\langle \epsilon_{\NHI,353} \rangle$ from the square patches, defined in Sect.\ \ref{sec:cib-ps-all}, at high Galactic latitudes. We applied the final iterative mask (discussed in Sect.\ \ref{sec:hmc-cib}) to the high-latitude square patches to select only the low \NHI pixels where the dust-\HI correlation holds. We also avoided square patches where more than $30\%$ pixels were masked. Finally, we had 40 such high-latitude square patches with a sufficient number of valid pixels for the correlation analysis. Using a simple linear regression with the \NHI map, we found a value of the emissivity per square patch. While performing linear regression, we took the standard deviation of the contamination as an error bar per pixel into account. By analysing 40 sky patches, we obtained the mean and standard deviation of the dust emissivity as $\langle \epsilon_{\HI,353} \rangle = 38\pm6\,\kJysr\paren*{\hiunit}$. We produced $25$ realisations of the $\ts_{\HI}$ maps by varying $\epsilon_{\HI,353}$ within the mean and standard deviation of this estimate. The bottom panel of Fig. \ref{fig:result-wph-data} shows the decomposition of $\ts$ in terms of the $\ts_{\HI}$ (left panel) and $\ts_{\molH}$ (middle panel) using the value of $\langle \epsilon_{\HI,353} \rangle$. Visually, the $\ts_{\molH}$ map is clumpy, whereas the $\ts_{\HI}$ map is more diffuse.
 
\begin{figure}[!htbp]
   \centering
   \includegraphics[width=\columnwidth]{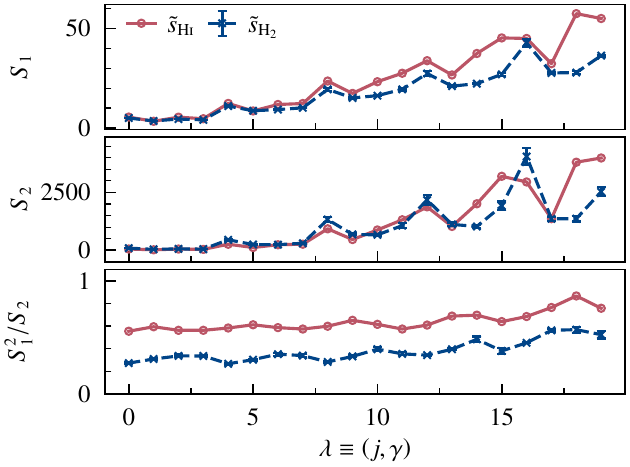}
      \caption{$S_{1}$, $S_{2}$, and $S_{1}^{2}/S_{2}$ for the $\ts_{\HI}$ map (red circles) and the $\ts_{\molH}$ maps (blue crosses with the dashed line). The standard deviation is obtained from $25$ realisations of $\ts_{\molH}$ maps.
      } 
         \label{fig:nhi-nh2-S1-S2}
   \end{figure}

 %%%%%%%%%%%%%%%%%%%%%%%%%%%%%%%%%%%%%%%%%%%%%%%%%%%%%%
We computed the SC statistics of the $\ts_{\HI}$ and $\ts_{\molH}$ maps to characterise their morphological differences. We used the $S_1$ and $S_2$ statistics and the $S_{1}^{2}/S_{2}$ ratio to measure sparsity \citep{lei2025apj}. Figure \ref{fig:nhi-nh2-S1-S2} shows the variation in $S_{1}$, $S_{2}$, and $S_{1}^{2}/S_{2}$ as a function of $\lambda\equiv\paren{j,\gamma}$. The error bars on the SC statistics of the $\ts_{\molH}$ map were computed from the $25$ realisations of $\ts_{\molH}$ maps. The first-order coefficient $S_{1}$ and wavelet power spectrum $S_{2}$ show that the $\ts_{\HI}$ map has more power at large angular scales and at all orientations than the $\ts_{\molH}$ map. The lower values of the ratio $S_{1}^{2}/S_{2}$ for $\ts_{\molH}$ compared to the $\ts_{\HI}$ map shows that $\ts_{\molH}$ map is significantly sparser than $\ts_{\HI}$, as the ratio decreases with increasing sparsity of the field. This is consistent with Fig.\ \ref{fig:result-wph-data}, where the localised high column density structures are only present in the $\ts_{\molH}$ map.
%%%%%%%%%%%%%%%%%%%%%%%%%%%%%%%%%%%%%%%%%%%%%%%%%%%%%
%%%%%%%%%%%%%%%%%%%%%%%%%%%%%%%%%%%%%%%%%%%%%%%%%%%%%%

Next, we computed the power spectrum of $\ts$ and compared it with the power spectrum of \NHI. Because of the limited sky coverage of the sky patch, the largest scale we obtained is $\ell \simeq 25$. We fitted them with a power-law model, $\Cl \propto \paren{\ell/75}^{\alpha}$, in the multipole range $25 \leqslant \ell \leqslant 625$ and found the best-fit value of the exponents as $\alpha_{\ts}= -2.51\pm0.07$ and $\alpha_{\HI}=-2.92 \pm 0.10$. The difference in the slope of $\ts$ and \NHI spectrum, ${\Delta \alpha = \alpha_{\ts} - \alpha_{\HI} \approx 0.4}$, also indicates a significant amount of dust emission associated with gas in the form of \molH~\citep{Desert:1988, Reach:1998}.
%%%%%%%%%%%%%%%%%%%%%%%%%%%%%%%%%%%%%%%%%%%%%%%%%%%%%%
%%%%%%%%%%%%%%%%%%%%%%%%%%%%%%%%%%%%%%%%%%%%%%%%%%%%%%
%%%%%%%%%%%%%%%%%%%%%%%%%%%%%%%%%%%%%%%%%%%%%%%%%%%%%%
%%%%%%%%%%%%%%%%%%%%%%%%%%%%%%%%%%%%%%%%%%%%%%%%%%%%%%
\subsection{Comparison of our dust map with the CSFD extinction map}\label{sec:csfd-vs-planck}

%%%%%%%%%%%%%%%%%%%%%%%%%%%%%%%%%%%%%%%%%%%%%%%%%%%%%%
In this section, we compare the recovered \Planck dust map at $353 \GHz$ with the CSFD map at $100\,\mum$ over the same sky patch as discussed in Sect.\ \ref{sec:data_results_maps}. The right column of Fig.\ \ref{fig:result-wph-data} shows the map-level comparison between $\ts$ (expressed in units of \kJysr) and $\scfd$ (expressed in units of $\rm mag$). Our first observation is that the \Planck dust map contains more localised structures than the CSFD map. This finding is supported by the power spectrum comparison between the two maps. We computed the power spectrum of $\ts$ and $\scfd$ over the apodised mask and fitted it with a power-law model within the multipole range $25 \leqslant \ell \leqslant 625$. The power spectrum exponent for the \Planck dust map is $\alpha_{\ts}=-2.51\pm0.07$ and that for the CSFD map is $\alpha_{\scfd}=-2.65\pm0.09$.

%%%%%%%%%%%%%%%%%%%%%%%%%%%%%%%%%%%%%%%%%%%%%%%%%%%%%%
Next, we computed the non-linear dependence of $\ts$ and $\scfd$ on the \NHI map by fitting a power-law model, $(\ts, \scfd) \propto \langle \NHI \rangle^{p}$. The power-law exponent $p$ is expected to be close to 1 in case of a tight correlation between dust and atomic gas. For each of the two dust maps, we grouped the pixel values into seven bins in ascending order of \NHI and computed the median values within each \NHI bin, as well as the 16 and 84 percentiles, which we used as error bars. We fitted the median values $\ts$ and $\scfd$ with the mean \NHI (taking the error bars into account) and found that the value of the exponent $p$ is $1.04$ for the \Planck map and $0.85$ for the CSFD map. This result shows that the slope of the power law differs significantly between the \Planck and CSFD maps. Furthermore, the data scatter in the \NHI correlation is greater for $\ts$ than for the $\scfd$ map. 

The morphological differences between the recovered \Planck dust map and CSFD map can be explained by the different spectral energy distribution (SED) of \HI-correlated dust emission and \molH-correlated dust emission. When the two SEDs follow the modified black-body spectrum (MBB), then the ratio of the emissivities of the \HI and \molH-correlated dust components at $353 \GHz$ can be written as
%%%%%%%%%%%%%%%%%%%%%%%%%%%%%%%%%%%%%%%%%%%%%%%%%%%%%%
\begin{equation}
    \frac{\epsilon_{\HI, 353}}{\epsilon_{\molH, 353}}  = \paren*{\frac{353}{\nu_0}}^{(\beta_{\HI}-\beta_{\molH})} \frac{B_{353}(T_{\HI} )}{B_{\nu_0}(T _{\HI})}\frac{B_{\nu_0}(T_{\molH} )}{B_{353}(T_{\molH})} \ ,\label{eq:em-ratio}
\end{equation}
%%%%%%%%%%%%%%%%%%%%%%%%%%%%%%%%%%%%%%%%%%%%%%%%%%%%%%
where $\nu_0=100\,\mu \rm m$ is the reference frequency, and $B_{353}\paren*{T}$ is the Planck black-body function at $353 \GHz$. Here, $(T_{\HI}, \beta_{\HI})$ and $(T_{\molH}, \beta_{\molH})$ are the temperatures and spectral indices  of the \HI and \molH-associated dust emission, respectively. We assumed that the ratio $\epsilon_{\HI, 100\,\mum}/\epsilon_{\molH,100\,\mum}=1$. When we neglected the attenuation of the interstellar radiation field by dust within the diffuse interstellar medium, the integral of the MBB spectrum or the dust radiance (given by Eq.~10 in \citealt{planck-XI:2013}) did not vary from \HI to the \molH gas. Under these two assumptions, the constraint relation between the SED parameters of  \HI and \molH-associated dust emission is
%%%%%%%%%%%%%%%%%%%%%%%%%%%%%%%%%%%%%%%%%%%%%%%%%%%%%%
\begin{equation}
    \frac{T^{(\beta_{\molH} +4)}_{\molH} \nu_0^{-\beta_{\molH}} \Gamma (\beta_{\molH} +4) \zeta (\beta_{\molH} +4)}{T^{(\beta_{\HI} +4)}_{\HI} \nu_0^{-\beta_{\HI}} \Gamma (\beta_{\HI} +4) \zeta (\beta_{\HI} +4)}=1 \ . 
\end{equation}
%%%%%%%%%%%%%%%%%%%%%%%%%%%%%%%%%%%%%%%%%%%%%%%%%%%%%%
We adopted $T_{\HI}=20 \,\rm K$ and $\beta_{\HI}=1.65$ from \cite{planck-XVII:2014}, computed from the MBB fit to the dust emissivities at the far-infrared frequencies $353$, $545$, and $857 \GHz$ and COBE-DIRBE $100\,\mum$ for the \HI-correlated dust emission. The change in $\beta_{\molH}$ follows from that of $T_{\molH}$, and thus, the change in relative grain emissivity at $353 \GHz$. We fitted the $\scfd$ map with two templates ($\ts_{\HI}$ and $\ts_{\molH}$) and a constant offset ($O$) over the valid pixels within the binary mask. We modelled the $\scfd$ map as
\begin{equation}
\begin{aligned}
    \scfd & = a_{\HI} \ts_{\HI} + a_{\molH} \ts_{\molH} + O \, , \label{eq:sim-nhi-nh2}\\
    & = a_{\HI} \epsilon_{\HI,353} \paren*{\NHI + \frac{a_{\molH} \ts_{\molH}}{a_{\HI} \epsilon_{\HI,353}}} + O\,,\\
    & = \epsilon_{\HI,100\,\mum}(\NHI + 2\NmolH) + O\,.
\end{aligned}
\end{equation}
%%%%%%%%%%%%%%%%%%%%%%%%%%%%%%%%%%%%%%%%%%%%%%%%%%%%%%
The best-fit values of $a_{\HI}$ is $0.32\pm0.06\,\rm mag (\MJysr)^{-1}$ and $a_{\molH}$ is $0.0321\pm0.0003\,\rm mag (\MJysr)^{-1}$. By defining the diffuse $\molH$ column density map ($\NmolH$) as $a_{\molH} \ts_{\molH}/(2 a_{\HI} \epsilon_{\HI,353})$, we enforced the emissivities of \HI and \molH-correlated dust emission to be the same at $100\,\mum$. With $\ts_{\molH} = \epsilon_{\molH,353} (2\NmolH)$, the ratio of the \HI-correlated dust emission and the \molH-correlated dust emission at $353 \GHz$ is obtained as $\epsilon_{\HI,353}/ \epsilon_{\molH,353}=0.11\pm 0.01$. The observed ratio can be achieved by assuming that the \molH-correlated dust emission has a temperature of $T_{\molH}=18.5\,\rm K$ and that the spectral index is $\beta_{\molH}=0.81$, which is different from the SED parameters of the \HI-correlated emission. Recently, \cite{Sullivan:2026} reported compelling evidence of two distinct dust emission components, which are referred to as hot and cold dust in the \Planck data, lending further support to our results.
%%%%%%%%%%%%%%%%%%%%%%%%%%%%%%%%%%%%%%%%%%%%%%%%%%%%%%
%%%%%%%%%%%%%%%%%%%%%%%%%%%%%%%%%%%%%%%%%%%%%%%%%%%%%% 
%%%%%%%%%%%%%%%%%%%%%%%%%%%%%%%%%%%%%%%%%%%%%%%%%%%%%
\section{Summary and discussion}\label{sec:discussion}

We statistically separated the dust signal from the CIB contamination at $353 \GHz$ using the SC statistics. We used the \NHI map as an external tracer to minimise the leakage of the contamination into the recovered dust signal map. To do this, the component-separation algorithm relied on a set of loss functions so that the recovered dust map retained the spatial correlation, with the \NHI map as present in the input data, and the contamination was statistically uncorrelated with the \NHI map. A brief summary and the main results of the analysis are given below.
%%%%%%%%%%%%%%%%%%%%%%%%%%%%%%%%%%%%%%%%%%%%%%%%%%%%%
\begin{itemize}
%%%%%%%%%%%%%%%%%%%%%%%%%%%%%%%%%%%%%%%%%%%%%%%%%%%%%%
\item We first analysed $25$ square patches of the contamination map obtained at low \NHI regions using the template-fit approach, taking the pixel-dependent dust emissivity and a global offset into account. The standard deviation of the contamination map and its angular power spectrum obtained from $25$ square patches are statistically consistent with each other. The three scalar MFs derived from these patches closely follow the expected values from a Gaussian distribution of CIB following the~\citet{Lenz:2019} best-fit CIB model plus instrumental noise contribution from FFP10 simulations. 
%%%%%%%%%%%%%%%%%%%%%%%%%%%%%%%%%%%%%%%%%%%%%%%%%%%%%%
\item We synthesised $300$ realisations of the contamination map using the scattering transform statistics based on the maximum entropy generative model. By construction, the $300$ synthesised contamination maps followed the same summary statistics as the $25$ contamination maps obtained from the template-fit approach at low \NHI regions.
%%%%%%%%%%%%%%%%%%%%%%%%%%%%%%%%%%%%%%%%%%%%%%%%%%%%%%
\item  We validated our algorithm on a simulated 2D test patch centred at Galactic coordinates $(l, b) = (45.0\degree, -54.3\degree)$. We used the interstellar dust-reddening map from~\cite{chiang2023apj} as a proxy of the dust emission and scaled its amplitude to different S/Ns with respect to the contamination map.  We added a synthesised realisation of the contamination to the simulated dust map. We computed the SC statistics of the input and recovered dust maps and the cross SC statistics of the residual map with \NHI to conclude that the component-separation algorithm works fairly well for $\mathrm{S/N}\geqslant 3$ case. There is no significant leakage of the dust emission into the contamination map. We verified that the recovered contamination map was uncorrelated with the \NHI map in the pixel space and in terms of $S_2$ statistics. We tested our component-separation algorithm on other sky patches and reached the same conclusions.
%%%%%%%%%%%%%%%%%%%%%%%%%%%%%%%%%%%%%%%%%%%%%%%%%%%%%%
\item We successfully applied the component-separation algorithm to a same test patch in the \Planck data ($\mathrm{S/N} \approx 8.4$) and separated $\ts$ from $\tr$. 
%%%%%%%%%%%%%%%%%%%%%%%%%%%%%%%%%%%%%%%%%%%%%%%%%%%%%%
\item We decomposed the component-separated \Planck dust map at $353 \GHz$ into a diffuse $\ts_{\HI}$ map and a clumpy $\ts_{\molH}$ map. The significant amount of dust associated with \molH gas in the $\ts$ can explain the difference in the power-law slopes of the $\ts$ and \NHI  angular power spectra. We used the SC statistics to understand the morphological difference between the two components. The $\ts_{\molH}$ map was found to be sparser but less filamentary in nature than the $\ts_{\HI}$ map.
%%%%%%%%%%%%%%%%%%%%%%%%%%%%%%%%%%%%%%%%%%%%%%%%%%%%%%
\item We also compared the component-separated dust map with the CSFD map at $100\,\mum$ for this region. The CSFD map is more tightly correlated with the \NHI map than the $\ts$ map. The maps and the difference in power spectra exponents show that the $\ts$ has more localised structures in it than the CSFD map. 
%%%%%%%%%%%%%%%%%%%%%%%%%%%%%%%%%%%%%%%%%%%%%%%%%%%%%
\end{itemize}
%%%%%%%%%%%%%%%%%%%%%%%%%%%%%%%%%%%%%%%%%%%%%%%%
This paper opens the way to producing dust-reddening maps using the \Planck data. The decomposition of the dust map (free from CIB contamination) into \NHI and \NmolH components helped us to investigate the formation of \molH gas within the diffuse ISM. In a future paper, we will analyse all the available \Planck high-frequency data ($\nu \geqslant 217$\, GHz) to model the component-separated dust emission in terms of a modified black-body spectrum. The next step is to expand the dust-CIB separation problem to encompass the full sky using the software package \texttt{Foscat}. The production of clean dust maps, devoid of any CIB contamination, at all HFI frequencies is crucial for determining the spectral energy distribution of dust emission and for validating the dust-\HI correlation in regions of low \HI column density regions.
%%%%%%%%%%%%%%%%%%%%%%%%%%%%%%%%%%%%%%%%%%%%%%%%%%%%%%
%%%%%%%%%%%%%%%%%%%%%%%%%%%%%%%%%%%%%%%%%%%%%%%%%%%%%%
\begin{acknowledgements} 

We thank Constant Auclair for the useful discussion and help during the start of the project. Some of the results in this paper have been derived using the \healpix package. We acknowledge use of the \Planck Legacy Archive. \Planck is an ESA science mission with instruments and contributions directly funded by ESA Member States, NASA, and Canada. The Parkes Radio Telescope is part of the Australia Telescope National Facility which is funded by the Australian Government for operation as a National Facility managed by CSIRO. The EBHIS data are based on observations performed with the 100-m telescope of the MPIfR at Effelsberg. EBHIS was funded by the Deutsche Forschungsgemeinschaft (DFG) under the grants KE757/7-1 to 7-3. The computations in this paper were run on the GPU cluster at NISER supported by the Department of Atomic Energy of the Government of India. S.\ S.\ is supported by the National Postdoctoral Fellowship of the Science and Engineering Research Board (SERB), ANRF, Government of India (File No.: PDF/2023/000594). This work received government funding managed by the French National Research Agency under France 2030, reference numbers ``ANR-23-IACL-0008'' and ``ANR-25-CE46-6634''. The authors thank the anonymous referee for the helpful suggestions and constructive comments that improved the quality and clarity of this paper.
\end{acknowledgements} 
%%%%%%%%%%%%%%%%%%%%%%%%%%%%%%%%%%%%%%%%%%%%%%%%%%%%%%
%%%%%%%%%%%%%%%%%%%% REFERENCES %%%%%%%%%%%%%%%%%%%%
\bibliographystyle{aa} % style aa.bst
\bibliography{sed} 
%%%%%%%%%%%%%%%%%%%%%%%%%%%%%%%%%%%%%%%%%%%%%%%%%%%%%%
\pagebreak
\newpage
%%%%%%%%%%%%%%%%%%%%%%%%%%%%%%%%%%%%%%%%%%%%%%%%%%%%%%
\appendix
\numberwithin{equation}{section}
\numberwithin{table}{section}
%%%%%%%%%%%%%%%%%%%%%%%%%%%%%%%%%%%%%%%%%%%%%%%%%%%%%%
\section{Emissivity maps}\label{app:emissivity-maps}

\begin{figure}[!hbtp]
\centering
\includegraphics[width=0.8\columnwidth]{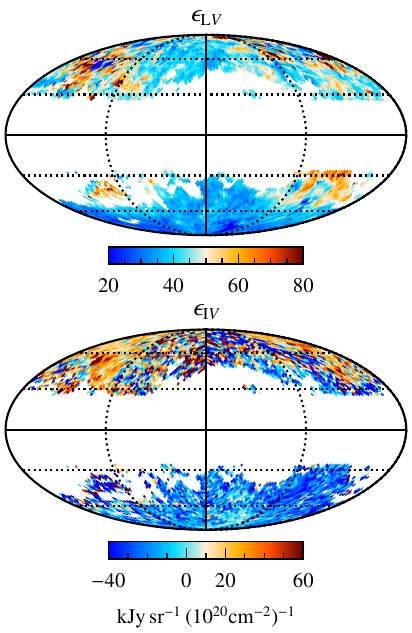}
\caption{LV-correlated dust emissivity map ($\epsilon_{\rm LV}$; \textit{top}) and IV-correlated dust emissivity map ($\epsilon_{\rm IV}$; \textit{bottom}) obtained from the template-fit approach as discussed in Sect.~\ref{sec:hmc-cib}. The $\epsilon_{\rm IV}$ map is mostly noisy in the southern Galactic hemisphere.}\label{fig:emissivity-maps}
\end{figure}

%%%%%%%%%%%%%%%%%%%%%%%%%%%%%%%%%%%%%%%%%%%%%%%%%%%%%%
Figure \ref{fig:emissivity-maps} shows the maps of LV-correlated and IV-correlated dust emissivity at \healpix $\Nside=32$ pixel resolution. The white areas represent pixels excluded from the template-fit analysis after applying the iterative mask.
%%%%%%%%%%%%%%%%%%%%%%%%%%%%%%%%%%%%%%%%%%%%%%%%%%%%%%
%%%%%%%%%%%%%%%%%%%%%%%%%%%%%%%%%%%%%%%%%%%%%%%%%%%%%%%
%%%%%%%%%%%%%%%%%%%%%%%%%%%%%%%%%%%%%%%%%%%%%%%%%%%%%%
%%%%%%%%%%%%%%%%%%%%%%%%%%%%%%%%%%%%%%%%%%%%%%%%%%%%%%
\section{Masks}\label{app:masks}
\begin{figure}[!htb]
\centering
\includegraphics[width=\columnwidth]{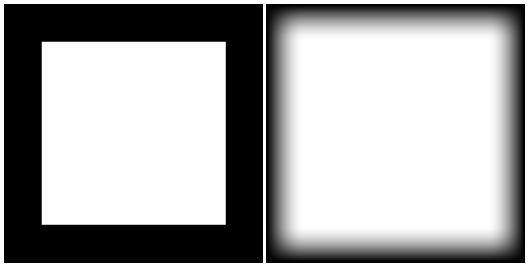}
\caption{Binary mask (\textit{left}) with inner $180\times 180$ pixels and the $256 \times 256$ pixels apodised mask (\textit{right}) used in our analysis.}\label{fig:square-mask}
\end{figure}

%%%%%%%%%%%%%%%%%%%%%%%%%%%%%%%%%%%%%%%%%%%%%%%%%%%%%
The binary and apodised masks used in this work are shown in Fig.\ \ref{fig:square-mask}. 
%%%%%%%%%%%%%%%%%%%%%%%%%%%%%%%%%%%%%%%%%%%%%%%%%%%%%
%%%%%%%%%%%%%%%%%%%%%%%%%%%%%%%%%%%%%%%%%%%%%%%%%%%%%%
%%%%%%%%%%%%%%%%%%%%%%%%%%%%%%%%%%%%%%%%%%%%%%%%%%%%%%
\section{Results for simulations}\label{app:sim-results-extra}

%%%%%%%%%%%%%%%%%%%%%%%%%%%%%%%%%%%%%%%%%%%%%%%%%%%%%
\subsection{Synthetic contamination maps}
%%%%%%%%%%%%%%%%%%%%%%%%%%%%%%%%%%%%%%%%%%%%%%%%%%%%%
\begin{figure} [!hbpt]
\centering
\includegraphics[width=\columnwidth]{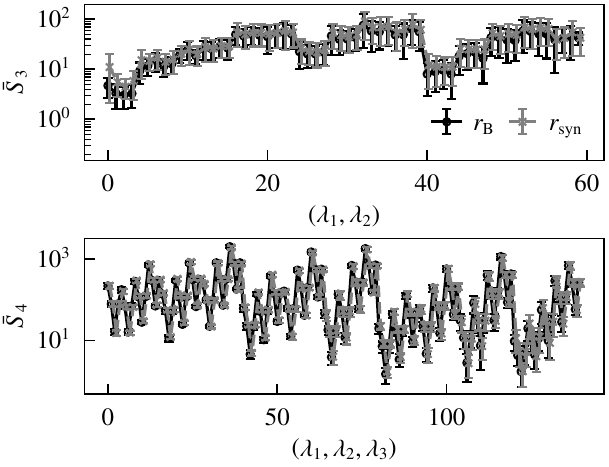}
\caption{Same as Fig.\ \ref{fig:hmc-wph-sc} but for $\bar{S}_{3}$ and $\bar{S}_{4}$ statistics. The coefficients are obtained by averaging over angles. The grey crosses are shifted in the $x$-axis.}
\label{fig:hmc-wph-sc-S3}
\end{figure}   
%%%%%%%%%%%%%%%%%%%%%%%%%%%%%%%%%%%%%%%%%%%%%%%%%%%%%
%%%%%%%%%%%%%%%%%%%%%%%%%%%%%%%%%%%%%%%%%%%%%%%%%%%%%
%%%%%%%%%%%%%%%%%%%%%%%%%%%%%%%%%%%%%%%%%%%%%%%%%%%%%%
In Fig.\ \ref{fig:hmc-wph-sc-S3} we present the normalised SC statistics ($\bar{S}_3$ and $\bar{S}_4$) for $25\,\rB$ regions and $300\,\rsyn$ maps, arranged in lexicographic order. As the contamination is independent of the wavelet orientations, the coefficients are averaged over angles. This demonstrates that the synthetic maps have identical SC statistics as the input contamination maps.
%%%%%%%%%%%%%%%%%%%%%%%%%%%%%%%%%%%%%%%%%%%%%%%%%%%%%
%%%%%%%%%%%%%%%%%%%%%%%%%%%%%%%%%%%%%%%%%%%%%%%%%%%%%%
\begin{figure}[!h]
\centering
\includegraphics[width=\columnwidth]{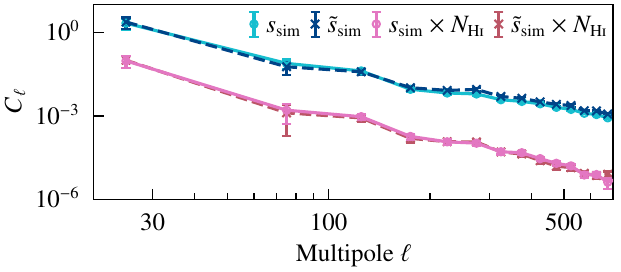}
\caption{Angular power spectra $\Cl$ vs.\ $\ell$ for input and output maps, showing auto-spectra of $\ssim$ (cyan circles) and $\tssim$ (blue crosses) and cross-spectra with the \NHI map (pink and red, respectively).}
\label{fig:sim-wph-dust-ps}
\end{figure}
%%%%%%%%%%%%%%%%%%%%%%%%%%%%%%%%%%%%%%%%%%%%%%%%%%%%%%
%%%%%%%%%%%%%%%%%%%%%%%%%%%%%%%%%%%%%%%%%%%%%%%%%%%%%
\subsection{Validating the component-separation algorithm}
%%%%%%%%%%%%%%%%%%%%%%%%%%%%%%%%%%%%%%%%%%%%%%%%%%%%%

Figure \ref{fig:sim-wph-dust-ps} shows the power spectra of the input and output dust maps. The spectra for $\ssim$ and $\tssim$ are consistent. We also successfully recovered the input correlation with the \NHI map.
%%%%%%%%%%%%%%%%%%%%%%%%%%%%%%%%%%%%%%%%%%%%%%%%%%%%%%
%%%%%%%%%%%%%%%%%%%%%%%%%%%%%%%%%%%%%%%%%%%%%%%%%%%%%
\begin{figure} 
\centering
\includegraphics[width=\columnwidth]{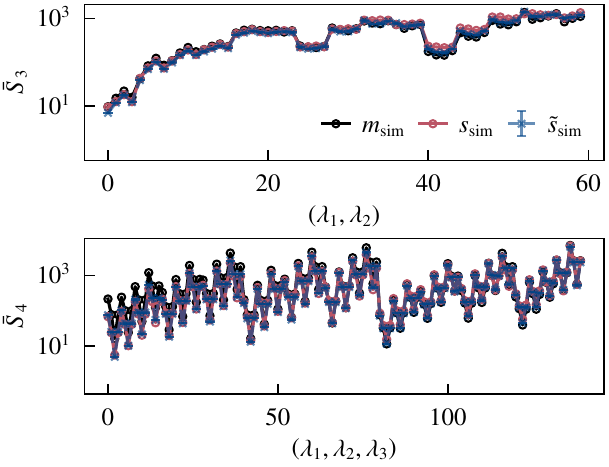}
\caption{Comparison of $\bar{S}_{3}$ and $\bar{S}_{4}$ statistics between $\msim$, $\ssim$ and $\tssim$. To reduce the number of coefficients and improve readability, we average over angles.}
\label{fig:sim-wph-sc-S3}
\end{figure}   
%%%%%%%%%%%%%%%%%%%%%%%%%%%%%%%%%%%%%%%%%%%%%%%%%%%%%

In Fig.\ \ref{fig:sim-wph-sc-S3} we show the normalised SC statistics, $\bar{S}_3$ and $\bar{S}_4$, for $\msim$, $\ssim$ and $\tssim$ in lexicographic order. The algorithm preserves the correlations between the different wavelet scales present in $\ssim$. Here, although the maps are anisotropic, we averaged over angles to reduce the number of coefficients and improve readability.
%%%%%%%%%%%%%%%%%%%%%%%%%%%%%%%%%%%%%%%%%%%%%%%%%%%%%

\begin{figure} [!hbpt]
\centering
\includegraphics[width=\columnwidth]{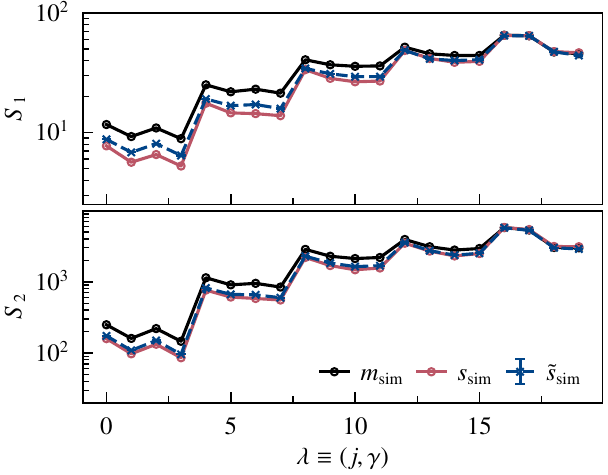}
\includegraphics[width=\columnwidth]{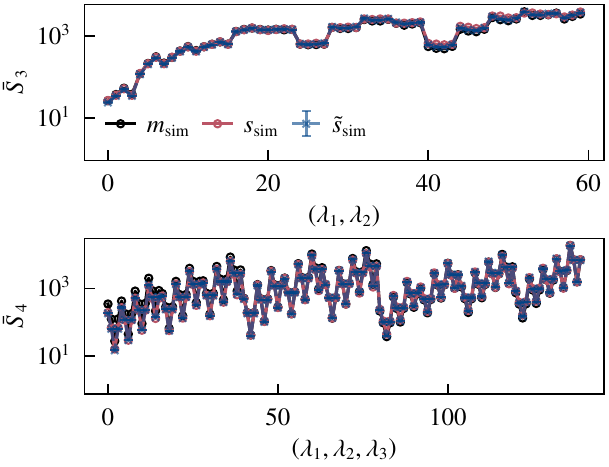}
\caption{Summary statistics $S_1$, $S_2$, $\bar{S}_{3}$, and $\bar{S}_{4}$ for the $\mathrm{S/N}=5$ simulation case.} 
\label{fig:sim-wph-sc-S3-snr5}
\end{figure}

Figure \ref{fig:sim-wph-sc-S3-snr5} shows the SC statistics ($S_1$, $S_2$, $\bar{S}_{3}$, and $\bar{S}_{4}$) for $\msim$, $\ssim$ and $\tssim$ at $\mathrm{S/N}=5$.
%%%%%%%%%%%%%%%%%%%%%%%%%%%%%%%%%%%%%%%%%%%%%%%%%%%%%
%%%%%%%%%%%%%%%%%%%%%%%%%%%%%%%%%%%%%%%%%%%%%%%%%%%%%
%%%%%%%%%%%%%%%%%%%%%%%%%%%%%%%%%%%%%%%%%%%%%%%%%%%%%
%%%%%%%%%%%%%%%%%%%%%%%%%%%%%%%%%%%%%%%%%%%%%%%%%%%%%
%%%%%%%%%%%%%%%%%%%%%%%%%%%%%%%%%%%%%%%%%%%%%%%%%%%%%
%%%%%%%%%%%%%%%%%%%%%%%%%%%%%%%%%%%%%%%%%%%%%%%%%%%%%
%%%%%%%%%%%%%%%%%%%%%%%%%%%%%%%%%%%%%%%%%%%%%%%%%%%%%%
\section{Statistics of the \Planck contamination map}\label{app:data-extra-temp}

\begin{figure}[!htbp]
\centering
\includegraphics[width=\linewidth]{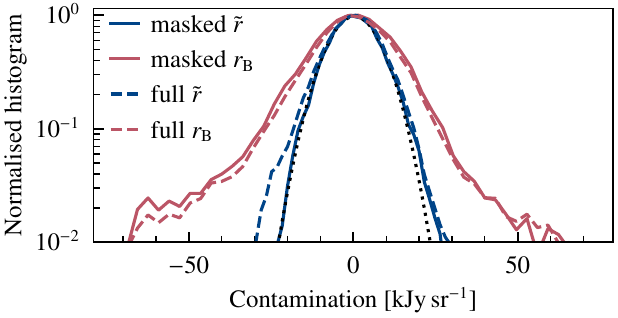}
\caption{1D distributions of $\tr$ (blue) and $\rB$ (red). The solid lines show pixels within the binary mask and the dashed lines within the full region. The dotted black line is the Gaussian fit to the distribution of $\tr$ over the binary mask. }
\label{fig:data-hmc-wph-hist}
\end{figure}

\begin{figure}[!hbtp]
\centering
\includegraphics[width=\columnwidth]{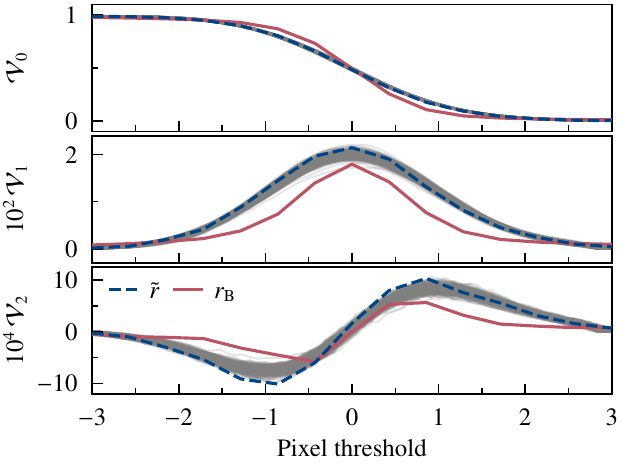}
\caption{ MFs of the $\tr$ (dashed blue line) and MFs of all $300$ realisations of $\rsyn$ (grey lines). The three MFs of $\tr$ match well with the corresponding MFs of $\rsyn$ used as an input to the component-separation algorithm. The solid red line represents the MFs of the $\rB$ map obtained from the template-fit approach.} 
\label{fig:data-wph-cib-mkf} 
\end{figure}

We compared the statistics of the recovered contamination map from SC-based component-separation method with the template-fit approach (discussed in Sect.~\ref{sec:hmc-cib}) over the same \Planck field as shown in Fig.\ \ref{fig:result-wph-data}. The histograms of the $\rB$ and $\tr$ are shown in Fig.\ \ref{fig:data-hmc-wph-hist} over the whole field and over the central unmasked region. From the histogram plot, it is clear that $\rB$ is highly non-Gaussian (independent of the mask) due to the presence of the Galactic residuals (dark gas, molecular $\molH$ and ionised hydrogen). The distribution of $\tr$ map is close to the Gaussian once we exclude the boundary pixels. The black dotted line in Fig.\ \ref{fig:data-hmc-wph-hist} is the Gaussian fit to the distribution of $\tr$ within the central unmasked region. Therefore, we can conclude that the SC-based component algorithm performs better than the template-fit approach in sky regions where excess \HI-uncorrelated Galactic emissions are present. 
%%%%%%%%%%%%%%%%%%%%%%%%%%%%%%%%%%%%%%%%%%%%%%%%%%%%%%
%%%%%%%%%%%%%%%%%%%%%%%%%%%%%%%%%%%%%%%%%%%%%%%%%%%%%
%%%%%%%%%%%%%%%%%%%%%%%%%%%%%%%%%%%%%%%%%%%%%%%%%%%%%
In Fig.\ \ref{fig:data-wph-cib-mkf} we show the MFs of $\tr$ (dashed blue line), $\rB$ (solid red line) and those from the $300\,\rsyn$ maps (solid grey lines) within the binary mask. The MFs of the $\tr$ map agree well with the $300\,\rsyn$ maps. Figures \ref{fig:data-hmc-wph-hist} and \ref{fig:data-wph-cib-mkf} demonstrate that the contamination map of the SC-based component-separation do not exhibit a non-Gaussian signature of Galactic emission.
%%%%%%%%%%%%%%%%%%%%%%%%%%%%%%%%%%%%%%%%%%%%%%%%%%%%%
%%%%%%%%%%%%%%%%%%%%%%%%%%%%%%%%%%%%%%%%%%%%%%%%%%%%%
%%%%%%%%%%%%%%%%%%%%%%%%%%%%%%%%%%%%%%%%%%%%%%%%%%%%%
%%%%%%%%%%%%%%%%%%%%%%%%%%%%%%%%%%%%%%%%%%%%%%%%%%%%%
\section{Comparison of the component-separated \Planck data results}\label{app:data-results-comp}

\begin{figure}[!htb]
\centering
\includegraphics[width=\columnwidth]{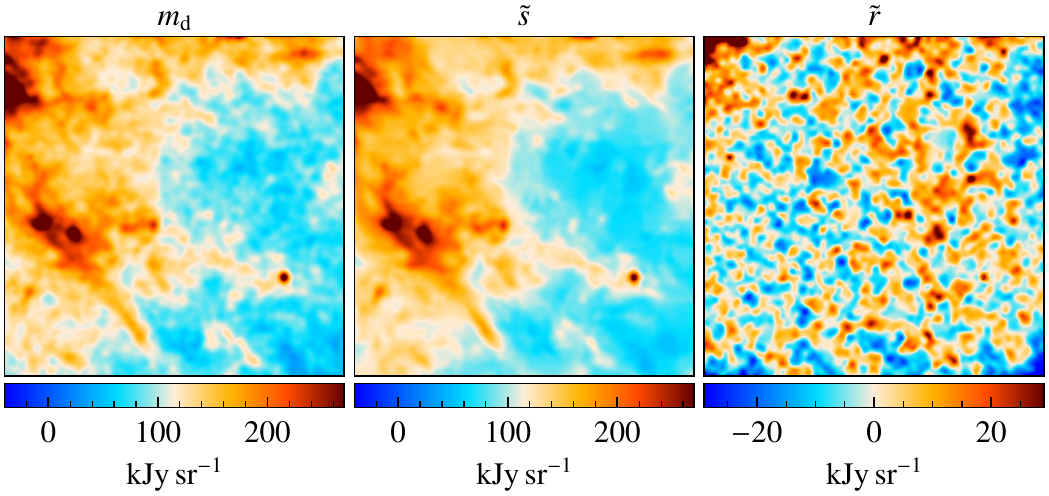}
\caption{\Planck data at $353 \GHz$ (\textit{left}),  $\ts$ (\textit{middle}), and $\tr$ (\textit{right}) for the same sky region as used in Sect.\ \ref{sec:planck-simulation}.} 
\label{fig:app-result-wph-data}
\end{figure}

%%%%%%%%%%%%%%%%%%%%%%%%%%%%%%%%%%%%%%%%%%%%%%%%%%%%%
%%%%%%%%%%%%%%%%%%%%%%%%%%%%%%%%%%%%%%%%%%%%%%%%%%%%%%
For completeness, we showed the results of the component-separation algorithm for the \Planck data region used to validate the component-separation algorithm in Sect.\ \ref{sec:planck-simulation}. Assuming the mean standard deviation of the contamination maps, $\sigma_{\rsyn}=9.0\,\kJysr$, this specific sky patch in the \Planck data has an $\mathrm{S/N}\approx 5$. The first column of Fig.\ \ref{fig:app-result-wph-data} shows $\md$, second column shows $\ts$, and third column shows $\tr$, highlighting the difference between the \Planck data and the recovered dust map. The first column of Fig.\ \ref{fig:app-result-wph-data} shows $\md$, second column shows $\ts$, and third column shows $\tr$, highlighting the difference between $\md$ and $\ts$. The standard deviation of $\tr$ computed over the binary mask is $8.3\,\kJysr$, consistent with the expected standard deviation of the $300\,\rsyn$ that are used as an input to the component-separation algorithm. 
%%%%%%%%%%%%%%%%%%%%%%%%%%%%%%%%%%%%%%%%%%%%%%%%%%%%%%
%%%%%%%%%%%%%%%%%%%%%%%%%%%%%%%%%%%%%%%%%%%%%%%%%%%%%%
%%%%%%%%%%%%%%%%%%%%%%%%%%%%%%%%%%%%%%%%%%%%%%%%%%%%%
\begin{figure}[!htb]
\centering
\includegraphics[width=\columnwidth]{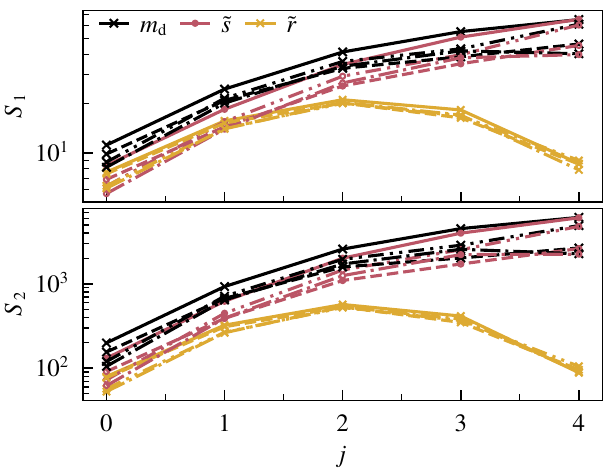}
\caption{$S_{1}$ and $S_{2}$ coefficients of the $\md$, $\ts,$ and $\tr$ maps. The four line types show the four different orientations.}
\label{fig:app-result-wph-S1-S2}         
\end{figure}
%%%%%%%%%%%%%%%%%%%%%%%%%%%%%%%%%%%%%%%%%%%%%%%%%%%%%%
%%%%%%%%%%%%%%%%%%%%%%%%%%%%%%%%%%%%%%%%%%%%%%%%%%%%%%%

The bright infrared point source centred at Galactic coordinates $(l, b) = (45.0\degree, -54.3\degree)$ is mostly present in the recovered dust map and a small fraction of it leaks into the residual map. We applied a $1\degree$ cut around the point source to exclude it from the further analysis of the summary statistics and the power spectrum. Figure\ \ref{fig:app-result-wph-S1-S2} shows the $S_{1}$ and $S_{2}$ coefficients for $\md$, $\ts$ and $\tr$ for the four different orientations of the wavelet. The $\md$ map at large scales (or high $j$ values) is signal dominated. The contamination map, which is statistically isotropic, has roughly the same power in all four orientations.
%%%%%%%%%%%%%%%%%%%%%%%%%%%%%%%%%%%%%%%%%%%%%%%%%%%%%%
\begin{figure}[!htb]
\centering
\includegraphics[width=\columnwidth]{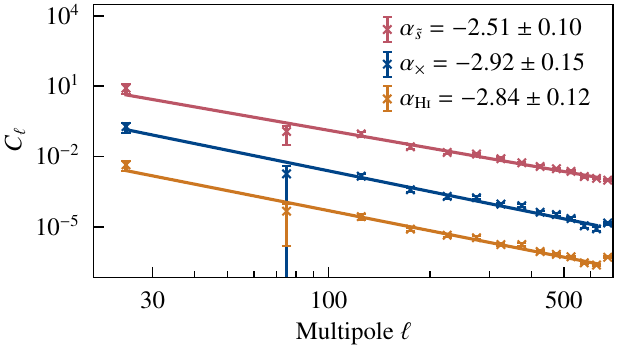}
\caption{Auto spectrum $\Cl$ with $\ell$ of $\ts$ (red crosses) and the best-fit power-law model (red line) in units of $\kJysrsq$. The blue shows the cross power spectrum of $\ts$ and the \NHI data along with the best-fit power-law model in $\kJysr\paren*{\hiunit}$. The orange shows the auto spectrum of the total \NHI data with the best-fit model in $\paren*{\hiunit}^{2}$.} 
\label{fig:app-result-wph-ps} 
\end{figure}
%%%%%%%%%%%%%%%%%%%%%%%%%%%%%%%%%%%%%%%%%%%%%%%%%%%%%%

Figure \ref{fig:app-result-wph-ps} shows the power spectra of $\ts$ and the \NHI data for the same region. We fitted the dust spectrum with the power-law model within the same multipole range as in Sect.\ \ref{sec:planck-data}. The straight lines in Fig.\ \ref{fig:app-result-wph-ps} show the best-fit power-law models for the auto-power spectrum of $\ts$, cross-power spectrum of $\ts$ with \NHI map, and auto-power spectrum of the \NHI map. The best-fit exponents are $\alpha_{\ts} = -2.51 \pm 0.10$, $\alpha_{\times}=-2.92 \pm 0.15$ and $\alpha_{\HI}=-2.84 \pm 0.12$, respectively. In this case, the difference in the slopes of $\ts$ and \NHI spectra, $\Delta \alpha = \alpha_{\ts} - \alpha_{\HI} \approx 0.3$. We inferred that $\Delta \alpha \ne 0$ could be due to dust emission associated with $\molH$.
%%%%%%%%%%%%%%%%%%%%%%%%%%%%%%%%%%%%%%%%%%%%%%%%%%%%%%
%%%%%%%%%%%%%%%%%%%%%%%%%%%%%%%%%%%%%%%%%%%%%%%%%%%%%
%%%%%%%%%%%%%%%%%%%%%%%%%%%%%%%%%%%%%%%%%%%%%%%%%%%%%
%%%%%%%%%%%%%%%%%%%%%%%%%%%%%%%%%%%%%%%%%%%%%%%%%%%%%%
\end{document}